\documentclass[preprint2,numberedappendix]{emulateapj}

\newcommand{\OII}{\ensuremath{{\rm [OII] \lambda 3726,3729} }}

\newcommand{\Msun}{\ensuremath{M_{\odot}}}

\newcommand{\Mstar}{\ensuremath{M_{\ast}}}
\newcommand{\Zsun}{\ensuremath{Z_{\odot}}}
\newcommand{\HST}{\emph{HST}}
\newcommand{\Spitzer}{\emph{Spitzer}}
\newcommand{\microm}{\ensuremath{{\rm \mu m } }}

\newcommand{\tenminusten}{$10^{-10}$ yr$^{-1}$}

\defcitealias{belli14lris}{Paper I}
\defcitealias{belli14mosfire}{Paper II}

\usepackage{capt-of}
\usepackage{graphicx}
\usepackage{subfigure}
\usepackage{verbatim}

\shorttitle{Stellar Populations from Spectroscopy of a Large Sample of $z>1$ Quiescent Galaxies}

\shortauthors{Belli, Newman and Ellis}
\submitted{Accepted for publication in the Astrophysical Journal}

\begin{document}

\title{Stellar populations from spectroscopy of a large sample of quiescent galaxies at \emph{z}$ > 1$: Measuring the contribution of progenitor bias to early size growth}

\author{Sirio Belli\altaffilmark{1}, Andrew B. Newman\altaffilmark{2,3}, Richard S. Ellis\altaffilmark{1}}
\altaffiltext{1}{Department of Astronomy, California Institute of Technology, MS 249-17, Pasadena, CA 91125, USA}
\altaffiltext{2}{The Observatories of the Carnegie Institution for Science, 813 Santa Barbara St., Pasadena, CA 91101, USA}
\altaffiltext{3}{Carnegie-Princeton Fellow}

\begin{abstract}

We analyze the stellar populations of a sample of 62 massive ($\log \Mstar/\Msun > 10.7$) galaxies in the redshift range $1 < z < 1.6$, with the main goal of investigating the role of recent quenching in the size growth of quiescent galaxies. We demonstrate that our sample is not biased toward bright, compact, or young galaxies, and thus is representative of the overall quiescent population. Our high signal-to-noise ratio Keck LRIS spectra probe the rest-frame Balmer break region which contains important absorption line diagnostics of recent star formation activity. We obtain improved measures of the various stellar population parameters, including the star-formation timescale $\tau$, age and dust extinction, by fitting templates jointly to both our spectroscopic and broad-band photometric data. We identify which quiescent galaxies were recently quenched and backtrack their individual evolving trajectories on the $UVJ$ color-color plane finding evidence for two distinct quenching routes. By using sizes measured in the previous paper of this series, we confirm that the largest galaxies are indeed among the youngest at a given redshift. This is consistent with some contribution to the apparent growth from recent arrivals, an effect often called \emph{progenitor bias}. However, we calculate that recently-quenched objects can only be responsible for about half the increase in average size of quiescent galaxies over a 1.5 Gyr period, corresponding to the redshift interval $1.25<z<2$. The remainder of the observed size evolution arises from a genuine growth of long-standing quiescent galaxies.

\end{abstract}


% ***************************************************************************************************
%						INTRODUCTION
% ***************************************************************************************************

\section{Introduction}
\label{sec:intro}

In the local universe, quiescent galaxies present a particularly tight \emph{red sequence} in the color-mass diagram \citep[e.g.,][]{bower92, blanton03, baldry04}. Understanding the mass assembly history of this remarkably homogeneous population remains one of the most important questions in the field of galaxy evolution. Quiescent galaxies selected at high redshift demonstrate that the red sequence seen locally was already in place at $z \sim 2$ \citep{cimatti04, labbe05,kriek08}. However, high redshift quiescent galaxies are significantly smaller at fixed stellar mass \citep[e.g.,][]{daddi05, trujillo06, vandokkum06, vandokkum08, cimatti08} raising the question of how such size growth occurred while maintaining the uniformity of the population. Although the inferred size evolution was initially questioned, subsequent studies have confirmed the result, ruling out biases in both the mass and size measurements at high redshift \citep[e.g.,][]{muzzin09,szomoru12}.

Among the physical processes that may be responsible for this surprising size growth, theoretical arguments favor minor mergers since they represent an efficient way to increase the size of a galaxy compared to the growth of its stellar mass \citep[e.g.,][]{naab09, hopkins10}. However, as the comoving number density of quiescent galaxies increases by about an order of magnitude between $z\sim2$ and $z\sim0$ \citep[e.g.,][]{muzzin13}, most of those observed locally cannot be the descendants of those at high redshift. The remainder were likely star-forming systems whose star  formation was quenched and subsequently arrived on the red sequence. As star-forming galaxies are typically larger than quiescent galaxies \citep[e.g.,][]{newman12}, some of the inferred growth with time in the quiescent population may be due to the later arrival of these quenched systems. It has been argued this effect, termed \emph{progenitor bias}, could explain part or all of the surprising size evolution in the quiescent population \citep[e.g.,][]{carollo13,poggianti13numberdensity}.

Detailed spectroscopic studies provide the most effective way to make progress in understanding both the physical origin of the size growth in quiescent objects as well as in disentangling the contribution from progenitor bias. In the first paper of this series \citep{belli14lris}, we investigated the size growth of quiescent galaxies to $z \sim 1.6$ using deep Keck LRIS spectroscopy of over 100 massive galaxies with $z>1$. We considered size evolution at fixed velocity dispersion arguing that the latter quantity should remain relatively constant with time even in the event of minor mergers \citep[e.g.,][]{hopkins09scalingrel}. By matching each high redshift galaxy to local samples with equivalent velocity dispersions, we concluded that physical size growth must have occurred and that progenitor bias alone cannot explain the observations. Moreover, the \emph{growth efficiency} $d \log R/ d \log M$ we inferred over $0<z<1.6$ is consistent with that expected for minor mergers, a conclusion in agreement with the frequency of likely associated pairs observed over this redshift interval in deep CANDELS data \citep{newman12}.

The present paper addresses the more challenging aspect of the observations. At redshifts above $z \sim 1.5$, the rate of size growth accelerates significantly. Specifically, in \citet{newman12} we found the growth at fixed stellar mass over $1.5<z<2.5$, an interval of only 2 Gyr,  is comparable to that which occurred in the subsequent 9 Gyr to the present epoch. However, in this redshift range, the frequency of likely minor mergers is insufficient to explain the rapid growth. To verify this remarkably rapid size growth, we recently extended our spectroscopic study to a smaller sample of $2<z<2.5$ quiescent galaxies using MOSFIRE, a new near-infrared multi-object spectrograph at the Keck observatory \citep{belli14mosfire}. Combining dispersion measures for this new sample with the limited number of similar $z>2$ measures in the literature \citep{vandokkum09,toft12,vandesande13} enabled us to measure the growth efficiency, which is too high to be consistent with the minor merging scenario. In addition to the shortage of observed associated pairs during this early period \citep{newman12}, numerical simulations in the $\Lambda$CDM framework are also unable to explain the fast growth rate \citep{nipoti12,cimatti12}. Given the fast rise in the comoving number density of quiescent systems, progenitor bias is likely to become more important at higher redshift, and is conceivably a significant factor in explaining growth in the $1.5 < z < 2.5$ interval.

A direct way to estimate the contribution of newly-quenched galaxies to the size growth of quiescent sources is to examine the size distribution as a function of the \emph{age} of the stellar population. This tests whether the most compact objects are the oldest as would be the case if the growth is mostly due to progenitor bias. Luminosity weighted stellar ages can be inferred from the detailed absorption features seen in the rest-frame optical spectra. However, as high quality spectra are required for accurate age measures, limited work has so far been possible at $z>1$ \citep[e.g.,][]{kriek06, kriek09, onodera12, whitaker13}.  The LRIS spectra of $1 < z < 1.6$ galaxies obtained for our velocity dispersion study \citep{belli14lris} are ideal for this purpose. In addition to being the largest systematic spectroscopic study of quiescent galaxies above $z\sim1$ to date, the rest-frame optical spectra include important features such as the Balmer absorption lines, the 4000\AA\ break, and the [OII] emission line, that are sensitive to the past star formation activity on various timescales that probe earlier activity out to $z\sim$2-2.5. As we will show in this paper, we can improve the age constraints by combining our spectroscopic measures with those derived from the spectral energy distributions derived over a wide wavelength range from publicly available multi-band photometric surveys. We undertake a comprehensive Bayesian analysis that takes into account simultaneously both photometric and spectroscopic data \citep[see][]{newman14}. This allows us to secure accurate stellar population parameters for a large representative sample of quiescent galaxies at $z>1$.

The main goal of the present work is therefore to study the {\emph size-age relation} for quiescent galaxies at $1 < z < 1.6$ and thereby to infer the past size evolution of the red sequence population, disentangling genuine physical growth of old sources from the contribution of newly-quenched sources (progenitor bias). Additionally, by reconstructing the past star formation of individual objects now observed on the red sequence, we can explore the mass assembly history and obtain new insights into the physical processes responsible for the quenching that transformed star-forming galaxies into passive objects.

An overview of the paper follows. In Section \ref{sec:data} we review the sample, briefly discussing the LRIS spectra and the auxiliary photometric data. In Section \ref{sec:physprop} we derive the stellar population properties by fitting templates to our Keck spectra, demonstrating the value of additional constraints that arise from the presence of  [OII] 3727 \AA\ emission. In Section \ref{sec:redsequence} we analyze in detail various components of the color-color diagram for our LRIS sample and use our stellar population parameters to reconstruct the past trajectories of individual quiescent galaxies, measuring how recently they arrived on the red sequence. This enables us to investigate the role of quenching in the observed size growth over $1.25 < z < 2$, and hence to quantify the effect of progenitor bias, in Section \ref{sec:size}. Finally, we summarize our results and discuss them in the context of galaxy quenching scenarios in Section \ref{sec:discussion}. Throughout this paper we use the AB magnitude system, and assume a $\Lambda$CDM cosmology with $\Omega_m = 0.3$, $\Omega_\Lambda = 0.7$, and $H_0 = 70$km s$^{-1}$ Mpc$^{-1}$.

% ***************************************************************************************************
%						DATA
% ***************************************************************************************************

\begin{figure*}[tbp]
\centering
\includegraphics[width=0.90\textwidth]{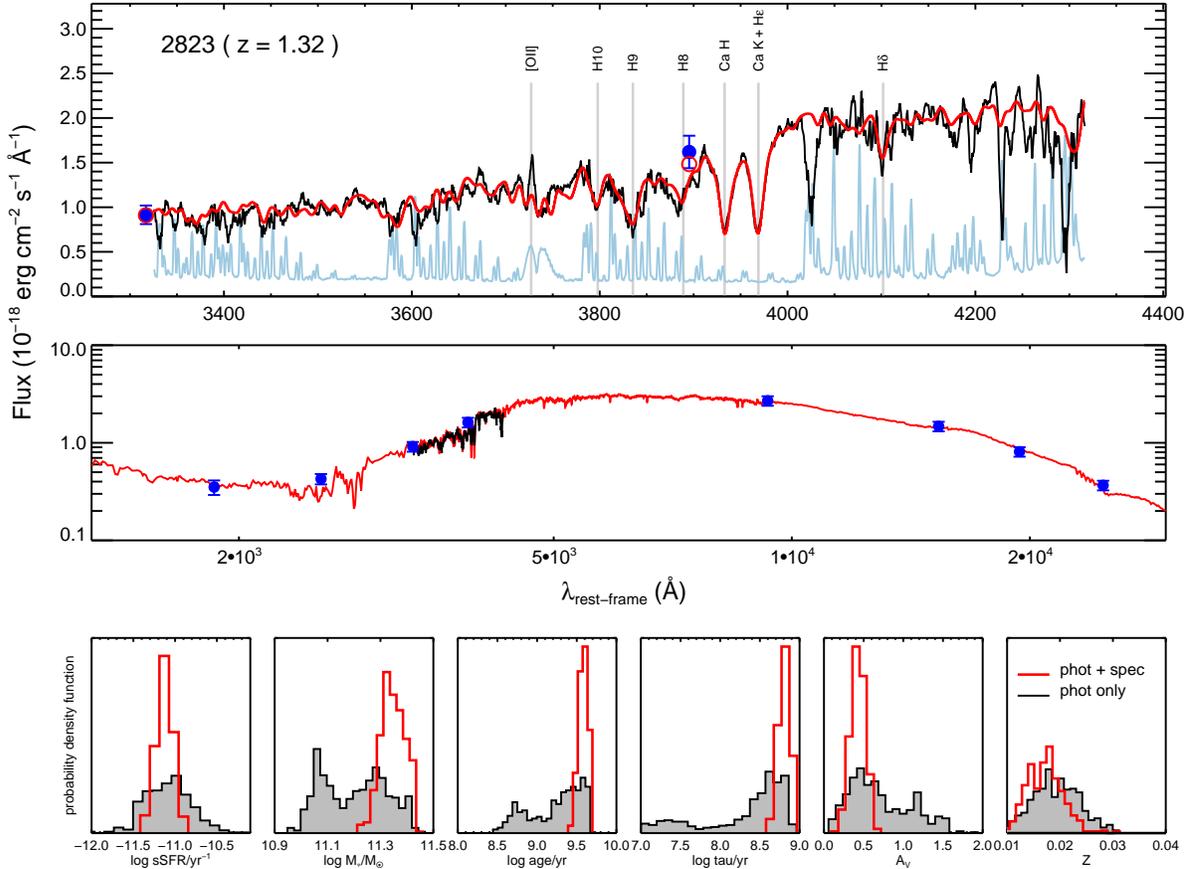}
\caption{An illustration of our spectral fitting technique for the object 2823 ($z=1.32$) which has a signal-to-noise ratio representative of the sample. Top two panels: observed Keck LRIS spectrum (black), error spectrum (cyan), observed multi-band photometry (blue) and best-fit model (red). In the top panel, empty red circles show the flux in the observed passbands expected from the best-fit model, and vertical gray lines mark the most important spectral features. Bottom: the posterior distributions output by {\tt pyspecfit} are shown for the five stellar population parameters and the specific star formation rate. Gray histograms represent those obtained by fitting the photometric data alone, while the red histograms show the same distribution when the LRIS spectrum is included.}
\label{fig:examplefit}
\end{figure*}

\section{Data}
\label{sec:data}

The present analysis is drawn from the spectroscopic sample of 103 galaxies presented in \citet{belli14lris}, hereafter \citetalias{belli14lris}, to which the reader is referred for a detailed description. In brief, most of the galaxies in the sample were selected to have photometric redshifts in the range $ 0.9 < z_\mathrm{phot} < 1.6 $ and stellar masses, derived from broad-band photometry, larger than $10^{10.7}\Msun$. Massive and quiescent objects were given a higher priority when designing the slitmasks. All targets were observed with the LRIS Spectrograph \citep{oke95} and its red-sensitive CCD on the Keck I telescope, with integration times ranging from 3 to 11 hours per mask. Examples of the LRIS spectra are shown in \citetalias{belli14lris}.

All except three galaxies in our sample lie in fields observed by the Cosmic Assembly Near-IR Deep Extragalactic Legacy Survey \citep[CANDELS,][]{grogin11,koekemoer11}. Therefore, high-quality \HST\ F160W observations, together with a wealth of broad-band photometric data, are publicly available. For each object, we collate space and ground-based observations from the near-UV to the near-infrared, including \Spitzer\ IRAC data \citep{cardamone10, whitaker11, bielby12, barro11, kajisawa11, newman10} and MIPS data from the \Spitzer\ archive.

In Appendix \ref{sec:completeness} we demonstrate that, for stellar masses above $10^{10.7} \Msun$, our sample is fully representative of the population of quiescent galaxies in this redshift range in terms of both colors and sizes. In the following analysis we will consider those 62 objects with stellar masses above this threshold. This remains the largest $z>1$ unbiased sample with high signal-to-noise spectra.

% ***************************************************************************************************
%						DERIVATION OF PHYSICAL PROPERTIES
% ***************************************************************************************************

\section{Derivation of Physical Properties}
\label{sec:physprop}

\subsection{Stellar Populations}
\label{sec:stellarpop}

Stellar population properties of high redshift galaxies are usually derived by model fitting of either broad-band photometry or a spectrum. Our LRIS spectra probe a rest-frame region rich in diagnostics of recent star formation activity, such as the Balmer lines and the 4000\AA\ break. Older stellar populations, however, contribute mainly to the near-infrared emission. To take advantage of both our high quality Keck spectra and the wealth of photometry available for our sample, we fit stellar population templates jointly to both the spectroscopic and photometric data. We use the Bayesian code {\tt pyspecfit} presented by \citet{newman14}, which performs a Markov Chain Monte Carlo sampling of the parameter space and outputs the posterior distributions, from which uncertainties and degeneracies can be properly estimated. 

We mask out the spectral region around [OII] emission and those pixels most contaminated by sky emission. We allow a polynomial correction to the observed spectrum in order to match the broad-band spectral energy distribution. We also add in quadrature a 5\% contribution to represent systematic errors to the photometry, and we exclude the IRAC datapoint at 8 \microm, which is susceptible to contamination by dust emission. In order to give appropriate weighting to the spectra and photometry, we run an initial fit that we use only to calculate the chi-square, which we then use to rescale the error spectra. 

We selected stellar population templates from the library of \citet{bruzual03}, and assume a \citet{chabrier03} initial mass function (IMF) and the \citet{calzetti00} dust extinction law. We adopt exponentially decreasing star formation histories (or \emph{$\tau$ models}), characterized by the age $t_0$ and timescale $\tau$ (with star formation rate proportional to $e^{-(t-t_0)/\tau}$), which have log-uniform priors in the range $ 10^8 \mathrm{yr} < t_0 < t_H$ and $10^7 \mathrm{yr} < \tau < 10^{10} \mathrm{yr}$, where $t_H$ is the age of the universe corresponding to the galaxy redshift, which is fixed to its spectroscopic value. The templates depend on two further parameters: the dust attenuation $A_V$, with the uniform prior $0 < A_V < 4$, and the metallicity $Z$, with a normal prior centered on the solar value $\Zsun=0.02$ and with a width of 0.005. The final output of the fitting procedure includes also the stellar mass \Mstar, obtained by scaling the best-fit template to the observed photometric data. The specific star formation rate (i.e., star formation rate per unit stellar mass) is not a free parameter, but is uniquely determined by the combination of $t_0$ and $\tau$.

Figure \ref{fig:examplefit} illustrates the procedure for a representative galaxy at $z=1.32$. The template provides an excellent fit to the observed photometry from the rest-frame UV to the near-infrared and also the detailed Keck spectrum. The fit is fully described by the five stellar population parameters $t_0$, $\tau$, $A_V$, $Z$, and \Mstar. The posterior distributions output by {\tt pyspecfit} for each parameter are shown in red in the bottom panels of Figure \ref{fig:examplefit}. The posterior distribution for the specific star formation rate, derived from the posteriors of $t_0$ and $\tau$, is also shown. In each panel, the posterior distribution obtained by fitting only the photometric data (but keeping the redshift fixed to its spectroscopic value) is shown as a gray histogram. The advantage of including the spectroscopic data in the fit is clear: the posterior distributions become much narrower. For example, the median uncertainty on the ages in our sample decreases from 0.10 dex to 0.05 dex when including the spectra. The stellar population parameters and their uncertainties are listed in Table \ref{tab:sample}.

\begin{figure}[tbp]
\centering
\includegraphics[width=0.45\textwidth]{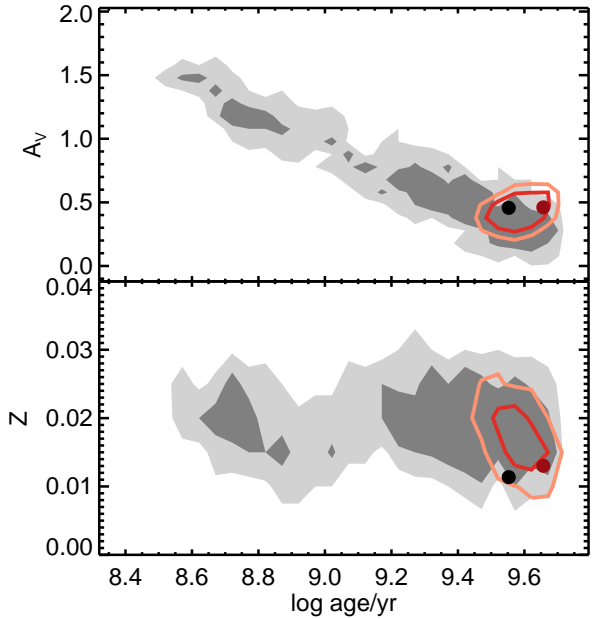}
\caption{2D posterior distributions for the object shown in Figure \ref{fig:examplefit}: dust extinction versus age (top panel) and stellar metallicity versus age (bottom panel). Grayscale contours represent the 68 and 95\% confidence levels for fit to the photometry alone with the black point marking the best-fit parameters. Red lines and points represent the fit to both the photometric and spectroscopic data. Combining both datasets is successful in breaking the dust-age degeneracy but less so for the metallicity-age degeneracy.}
\label{fig:degeneracy}
\end{figure}

Including the rest-frame spectra in the fitting procedure breaks degeneracies between some of the stellar population parameters. A familiar degeneracy is that between age and dust extinction, each of which has a similar reddening effect on the spectral energy distribution. The Balmer absorption lines and other features in the rest-frame optical spectrum, marked in Figure \ref{fig:examplefit}, are only sensitive to the age. Once the age is well determined spectroscopically, the amount of dust extinction is much more effectively constrained. The top panel of Figure \ref{fig:degeneracy} shows the two-dimensional (2D) posterior distribution of dust extinction and age for the galaxy presented in Figure \ref{fig:examplefit} and how inclusion of the spectrum improved estimates of both. A further degeneracy is that between age and metallicity, for which the 2D posterior distribution is shown in the bottom panel. In this case our technique is somewhat less successful.

The fitting procedure usually yields posterior distributions that are smooth and well separated from the edges of the prior. In only three cases (that we will discuss in Section \ref{sec:redsequence_diversity}) the age and $\tau$ parameters have the minimum allowed values. We discard these objects from our sample, since their star formation histories are unreliable. Broadly speaking the uncertainties in each parameter are comparable with those given in the example in Figure \ref{fig:examplefit} and this is important to remember in the following section.

\begin{figure}[tbp]
\centering
\includegraphics[width=0.5\textwidth]{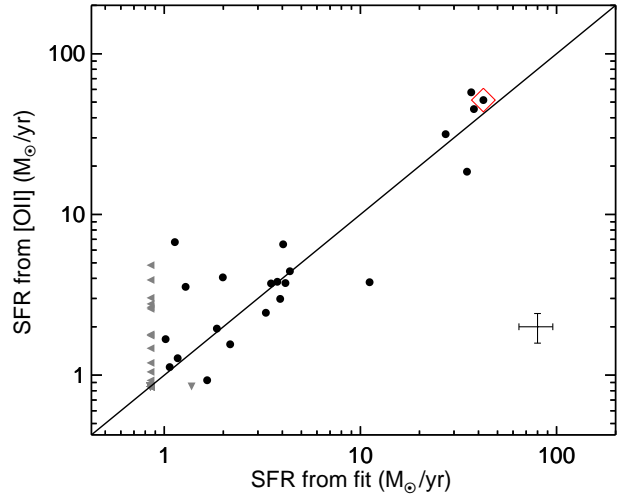}
\caption{A comparison of the star formation rate as derived from our spectral fitting technique with that estimated from the strength of [OII] 3727 \AA\ emission. Median uncertainties are shown in the bottom right corner, and upper limits are marked as gray triangles. Objects for which the IRAC colors imply the presence of an AGN are marked with red diamonds.}
\label{fig:sfr_oii}
\end{figure}

\begin{figure*}[htbp]
\centering
\includegraphics[width=\textwidth]{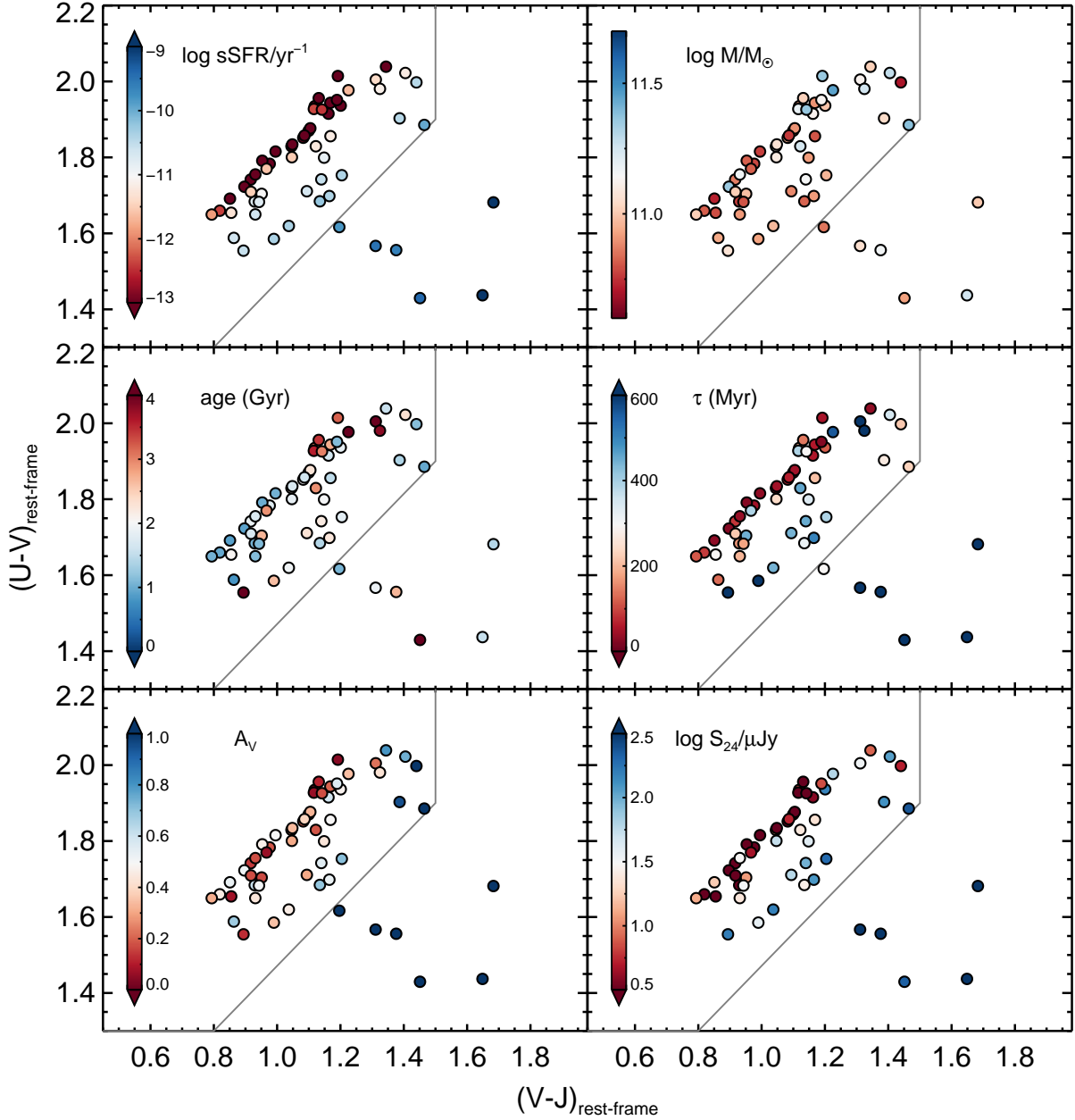}
\caption{The distribution on the $UVJ$ plane of the physical properties of the LRIS sample. Panels show the stellar population parameters obtained via spectral fitting for each galaxy, and the 24 \micron\ flux. The gray line indicates the division between quiescent and star-forming galaxies adopted by \citet{muzzin13}. In the last panel, only galaxies with available MIPS data are shown.}
\label{fig:uvj}
\end{figure*}

\subsection{[OII] Emission}
\label{sec:oii}

Many galaxies in our sample show \OII\ emission which is useful as an additional diagnostic of the current star formation rate, independent of the fitting procedure described above. Accordingly, we measured the [OII] rest-frame equivalent width for each spectrum by first subtracting the best-fit model spectrum from the observed one and fitting a double Gaussian to the residual. Both components of the [OII] doublet were assumed to have a fixed relative wavelength and identical width. Line emission with an equivalent width larger than 2\AA\ is seen for 40 out of 58 objects for which the observed spectra cover the appropriate wavelength range.

To derive star formation rates, we convert the equivalent widths to luminosities using the continuum flux given by our best-fit model spectra. We use the \citet{kewley04} calibration to derive the star formation rate, which we then correct for dust extinction. Figure \ref{fig:sfr_oii} compares the star formation rates obtained in this way with those obtained via spectral fitting. For galaxies with a significant level of star formation (i.e., above $\sim$1 \Msun/yr), the spectral fitting star formation rates are in good agreement with the ones derived from [OII] emission. Although we do not use the star formation rates in our main analysis, the agreement between the two estimates represents an important independent confirmation of the stellar population parameters obtained with {\tt pyspecfit}. 

A number of galaxies that are not forming stars according to the spectral fitting show weak, but clearly detected emission lines. Other than star formation, possible causes for the presence of an [OII] line are AGN and LINER emission. We use IRAC colors to identify strong AGNs, following \citet{donley12}, and find only two. Both are star-forming objects, and one has [OII] in the observed range and is marked with a red diamond in Figure \ref{fig:sfr_oii}. The [OII] lines detected in quiescent galaxies are therefore due to LINER emission, in agreement with what found at $z\sim0$ \citep{yan06,graves07} and $z\sim1$ \citep{lemaux10}. Such emission might be caused by hot old stars and is not necessarily associated with AGNs \citep{singh13}. 

In the subsequent analysis, we exclude the two strong AGNs from our sample. We also checked the X-ray emission using Chandra data, and found four detections in addition to the two strong AGNs (also detected). As these targets do not show any peculiarity, we keep them in our sample.

% ***************************************************************************************************
%						RED SEQUENCE
% ***************************************************************************************************

\section{The Red Sequence}
\label{sec:redsequence}

As discussed in Paper I, quiescent galaxies in our LRIS sample can be identified using a $UVJ$ color-color diagram \citep[e.g.,][]{wuyts07,williams09}. Figure \ref{fig:uvj} shows how the stellar population parameters obtained via spectral fitting (as described in Section \ref{sec:physprop}) are distributed according to the location of the galaxy in this diagram (see Appendix \ref{sec:completeness} for details on the rest-frame colors). In each panel the solid line indicates the division between quiescent and star-forming galaxies adopted by \citet{muzzin13}. 

Even in the redshift range $1<z<1.6$ a familiar picture emerges. A tight red sequence is clearly visible with a sharp upper envelope. Red sequence galaxies have low specific star formation rates, mature ages and relatively short $\tau$ parameters. Moreover, they have little to no dust extinction. Elsewhere in the diagram, `blue cloud' galaxies present significant star formation rates and dust extinction with larger $\tau$ parameters.

The last panel of Figure \ref{fig:uvj} show the distribution of the \Spitzer\ MIPS 24 \micron\ flux. As with the earlier discussion of \OII\ emission, this measure is completely independent of the spectral fit and supports the above picture. In particular, we note that the objects that comprise the tightest part of the red sequence have very low mid-infrared emission. Clearly they are genuinely quiescent galaxies and their red colors are not due to the effect of dust extinction.

\subsection{Diversity among Quiescent Galaxies}
\label{sec:redsequence_diversity}

\begin{figure}[tbp]
\centering
\includegraphics[width=0.5\textwidth]{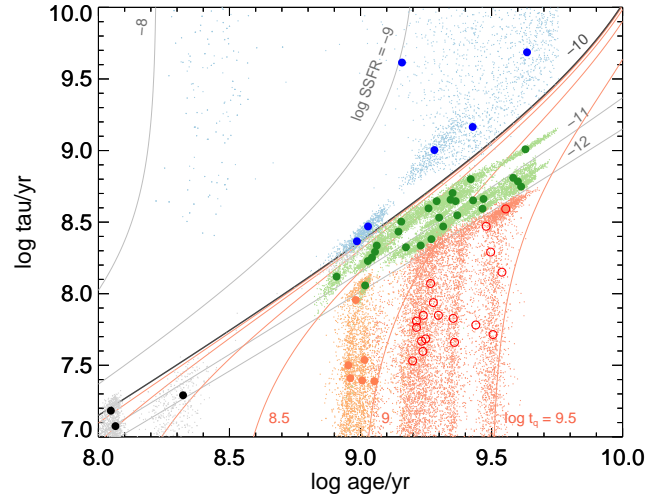}
\caption{Distribution of the stellar population parameters $\tau$ and age obtained via our fitting technique (Section \ref{sec:stellarpop}). Large points indicate the best-fit values, and the posterior distributions are plotted using small dots. The colors represent different galaxy populations: blue cloud (blue), green valley (green), red sequence (red, open symbols) and post-starburst galaxies (orange) - see text for definitions. The objects shown in black have posterior distributions limited by the prior boundaries, and we consider these to be less reliable. The gray lines mark regions of the plot of constant specific star formation rate, while the red lines mark regions of constant quiescent time (as defined in Section \ref{sec:size-age}).}
\label{fig:tau-age}
\end{figure}

Our high quality spectra allow us to go beyond the simple division of the population into star-forming and quiescent galaxies that is conventionally done at high redshift. Thus we depart briefly from our goal of analyzing the nature of size evolution of the quiescent population to illustrate this surprising diversity in the quiescent population. From Figure \ref{fig:uvj} we see that \emph{perpendicular} to the red sequence, the star formation rate increases progressively. Objects with intermediate values of specific star formation rate are often considered to be transitional objects moving toward the red sequence, particularly at high redshift \citep[e.g.,][]{goncalves12}. This population shows similar ages to the red sequence, but larger $\tau$ values, consistent with elevated levels of star formation. 

\begin{figure*}[tbp]
\centering
\includegraphics[width=0.8\textwidth]{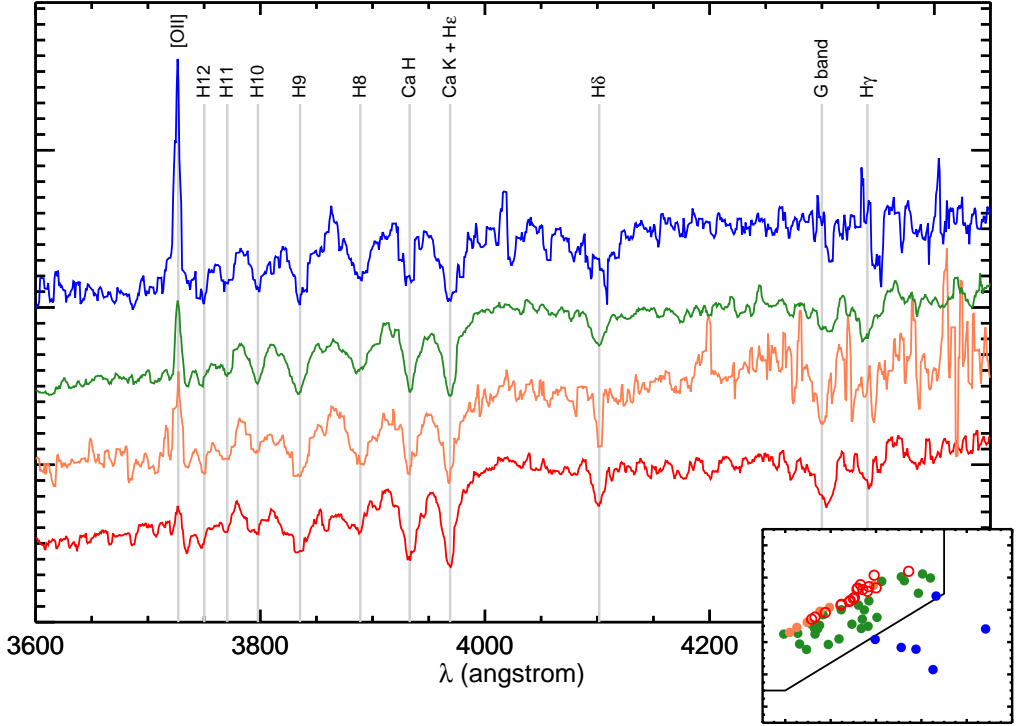}
\caption{Stacked spectra for the four galaxy populations defined in Section \ref{sec:redsequence_diversity}. Gray vertical lines mark the location of important spectral features. The inset shows the distribution of the populations on the $UVJ$ diagram. Colors and symbols as in Figure \ref{fig:tau-age}.}
\label{fig:stacks}
\end{figure*}

On closer examination, our stellar population parameters indicate there is some diversity even within the red sequence population itself. Figure \ref{fig:uvj} shows there is a clear gradient in the age along the sequence, from $\sim$1 Gyr at the blue end to $\sim$3 Gyr at the red end. The redder galaxies also tend to be more massive and less dusty. To better understand how this diversity might arise, we consider their distribution in the $\tau$ vs age plane in Figure \ref{fig:tau-age}. For each object we plot the best-fit value as a large data point and the full posterior distribution with small dots which is helpful in indicating the uncertainties. Lines of constant specific star formation rate are indicated. We identify different galaxy populations in Figure \ref{fig:tau-age}:
\begin{itemize}
\item Galaxies above the bold line, which marks a specific star formation rate of \tenminusten, are \emph{star-forming} (blue points, 6 objects). As a reference, the main sequence at this redshift corresponds to a specific star formation rate of $10^{-9}$ yr$^{-1}$ \citep{speagle14}.
\end{itemize}
We call \emph{quiescent} all the galaxies below the bold line. We adopt the value \tenminusten\ because it is roughly equivalent to a mass doubling time twice the age of the universe at $z\sim1.3$. This threshold in specific star formation rate is almost exactly equivalent to the $UVJ$ selection box shown in Figure \ref{fig:uvj}. We further divide quiescent galaxies into three groups:
\begin{itemize}
\item \emph{Green valley} galaxies are defined as having a specific star formation rate between $10^{-12}$ and $10^{-10}$ yr$^{-1}$ (green points, 27 objects). The posterior distributions of these galaxies are elongated, following lines of constant specific star formation rate. This indicates that the measurement of star formation activity is robust but there is a small degeneracy between age and $\tau$.
\item The \emph{red sequence} consists of genuinely old, passive galaxies, with ages above 1.25 Gyr and specific star formation rates below $10^{-12}$ yr$^{-1}$ (red empty points, 18 objects). The posterior distributions are vertical: for these objects we have a good measure of the age but only an upper limit on $\tau$, and therefore we can only obtain an upper limit on the star formation rate.
\item The remaining passive galaxies, i.e. those with ages below 1.25 Gyr and specific star formation rates below $10^{-12}$ yr$^{-1}$, are \emph{post-starburst galaxies} (orange points, 6 objects). We use this term to indicate quiescent galaxies that show signs of recent star formation activity; this is different from the often used definition in terms of absence of [OII] emission and presence of strong Balmer absorption lines \citep[e.g.,][]{dressler13}.
\end{itemize}
Finally, the points in black represent three galaxies whose determined $\tau$ and age are unphysically small and represent limits governed only by the boundary of the priors. We discard these objects from our analysis of the quiescent sample, since their colors are clearly in the star-forming region of the $UVJ$ diagram.

A striking way to further visualize this diversity in the population of quiescent galaxies is via stacked spectra for the four populations (Figure \ref{fig:stacks}) defined above. For each galaxy within the relevant population, we convolve the spectrum with a Gaussian kernel to yield a fixed velocity dispersion of 400 km s$^{-1}$, normalize to a median flux at 4000\AA$ < \lambda < 4050$\AA, and produce a median-stack. No weighting is applied to avoid biasing the results to more luminous objects. The spectra show a very clear decline in activity from blue cloud to old red sequence sources as indicated in a declining level of [OII] emission but an increasing 4000\AA\ break, more prominent G band and deep Calcium absorption lines, the latter being features associated with older stars. Importantly, however, these trends continue within the red sequence itself from the younger end (populated by post-starburst galaxies) to the older end. In the inset of Figure \ref{fig:stacks} we plot the four populations on the $UVJ$ diagram. Clearly the post-starburst galaxies occupy the blue side of the red sequence \citep[e.g.,][]{whitaker13}.

The purpose of this interlude in our goal to address size evolution will become clearer when we attempt to physically understand how these various subsets of quiescent galaxies fit into an evolutionary picture in Section \ref{sec:discussion}.

\subsection{Reconstructing the Quiescent Population}
\label{sec:redseq_evo}

The availability of ages and $\tau$ parameters for each LRIS galaxy in Figure \ref{fig:uvj} enables us to reconstruct their past star formation histories and hence their earlier trajectories on the $UVJ$ color-color diagram. This provides the basic means by which we can disentangle which quiescent sources are truly old and possibly growing in physical size, and which sources became quiescent more recently and may contribute to apparent growth with time via progenitor bias. 

\begin{figure}[tbp]
\centering
\includegraphics[width=0.5\textwidth]{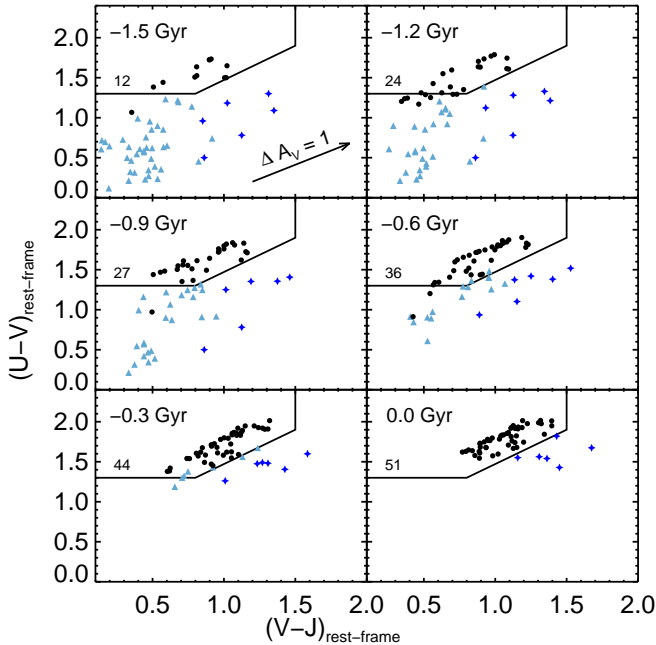}
\caption{Reconstructed evolution of the LRIS sample on the $UVJ$ diagram in a series of time snapshots 300 Myr apart up to the epoch of observation in the final panel (corresponding to the median redshift of our sample: $z=1.25$). For each time snapshot, black points represent quiescent galaxies, while light blue triangles are star-forming galaxies that will become quiescent by the end of the simulated evolution (i.e., at the time of observations). Blue stars represent galaxies that are star-forming throughout the simulation. The number of quiescent galaxies, defined as those with a specific star formation rate less than \tenminusten (Section \ref{sec:redsequence_diversity}), is 
shown in each panel.}
\label{fig:redseq_evo}
\end{figure}

We use the star formation history to calculate the stellar population parameters, including the rest-frame colors, at various periods earlier in time. In Figure \ref{fig:redseq_evo} we plot the distribution for the epoch of observation, $t_\mathrm{obs}$ (final panel), and at five earlier times $t_\mathrm{obs} - \Delta t$, with $\Delta t$ in increments of 300 Myr. These panels show clearly how the currently-observed red sequence of LRIS galaxies assembled over the previous 1.5 Gyr. At each time snapshot, we define galaxies with specific star formation rate under \tenminusten\ as quiescent, and we show them with black points in Figure \ref{fig:redseq_evo}. The reason we prefer to make this definition in terms of the specific star formation rate as opposed to directly selecting quiescent sources from the $UVJ$ diagram is that in calculating the evolutionary tracks we must assume that dust content and metallicity do not evolve. Since star-forming galaxies are observed to be on average more dust-rich than quiescent galaxies, quenching must to some extent also be associated with a decline in extinction. This means that our predicted past colors will generally be too blue for those galaxies that are quiescent at $\Delta t=0$, but that are still forming stars at earlier epochs (shown as blue triangles in Figure \ref{fig:redseq_evo}). The effect of dust extinction is shown by the arrow in the first panel. Clearly, a reasonable amount of dust can shift the population of transitional objects and bring it closer to the green valley, where galaxies are observed to lie. In the figure we also show the reconstructed evolution for the sample of 6 star-forming galaxies. However, we do not include these objects in the subsequent analysis as this sample is small and biased toward bright objects, unlike our quiescent sample.

\begin{figure}[tbp]
\centering
\includegraphics[width=0.5\textwidth]{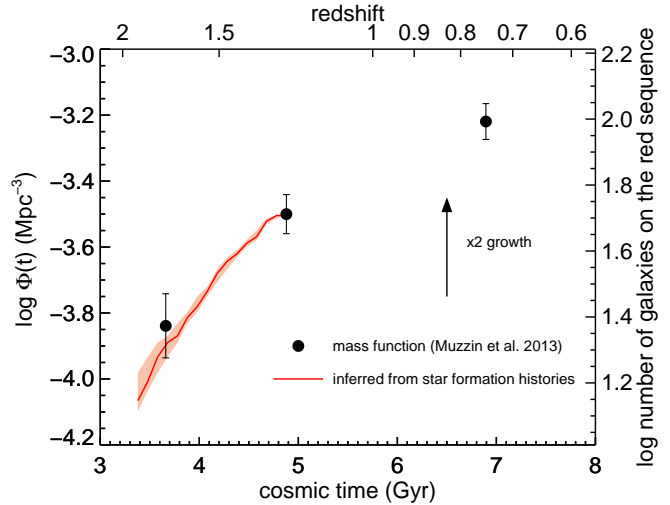}
\caption{The evolving number density of quiescent galaxies with $\log \Mstar/\Msun > 10.7$ from the stellar mass function study of \citet[][black points]{muzzin13} with respect to the left ordinate axis. The red line represents the evolution inferred from the star formation history analysis of our LRIS spectroscopic sample of quiescent galaxies whose median redshift is $z=1.25$ with respect to the right ordinate axis. The shaded area shows the effect of the uncertainties on the star formation histories. The vertical offset between the two samples is arbitrary given the uncertain volume probed by our spectroscopic survey.}
\label{fig:redseq_growth_compare}
\end{figure}

\begin{figure*}[tbp]
\centering
\includegraphics[width=0.45\textwidth]{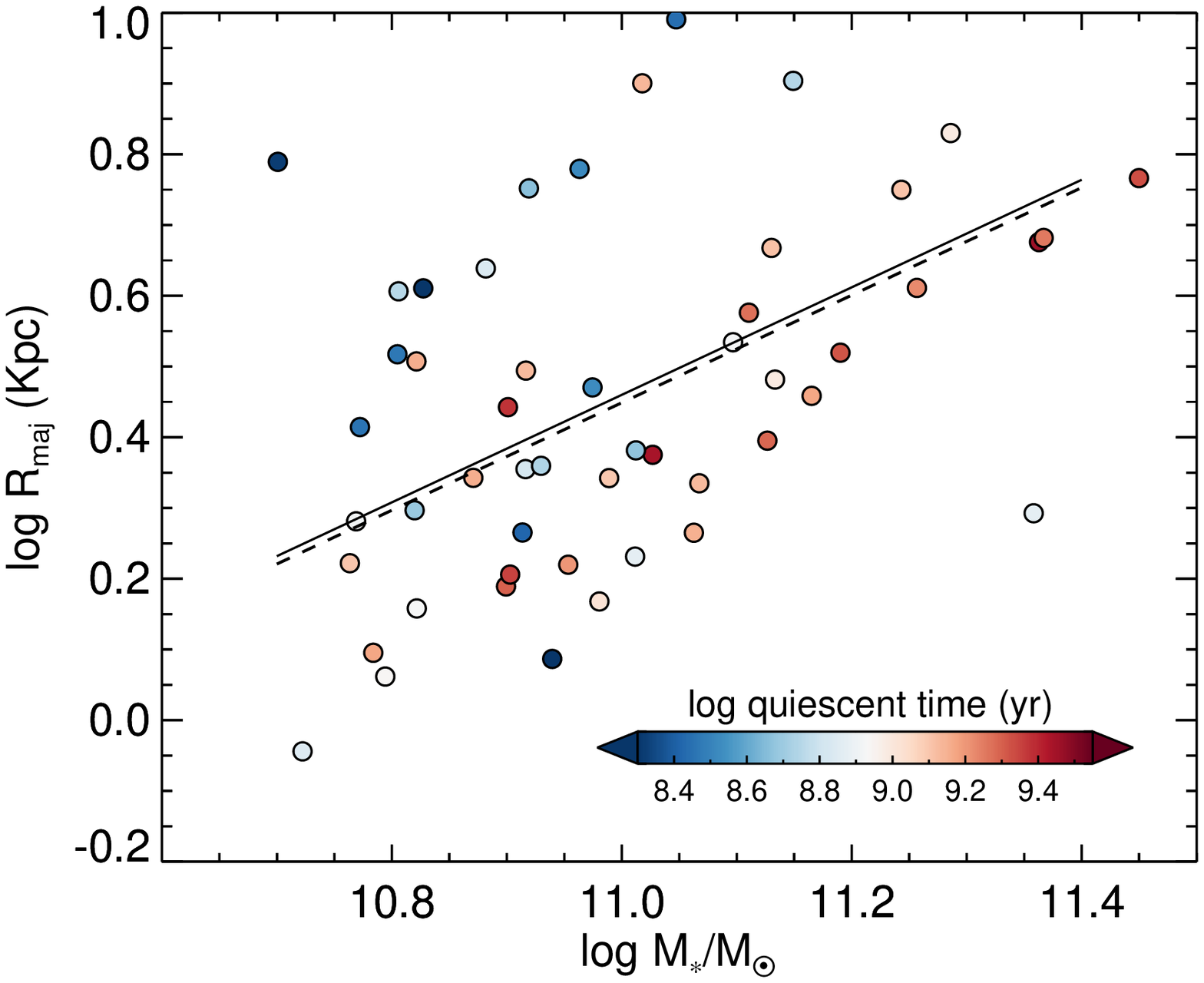}
\includegraphics[width=0.45\textwidth]{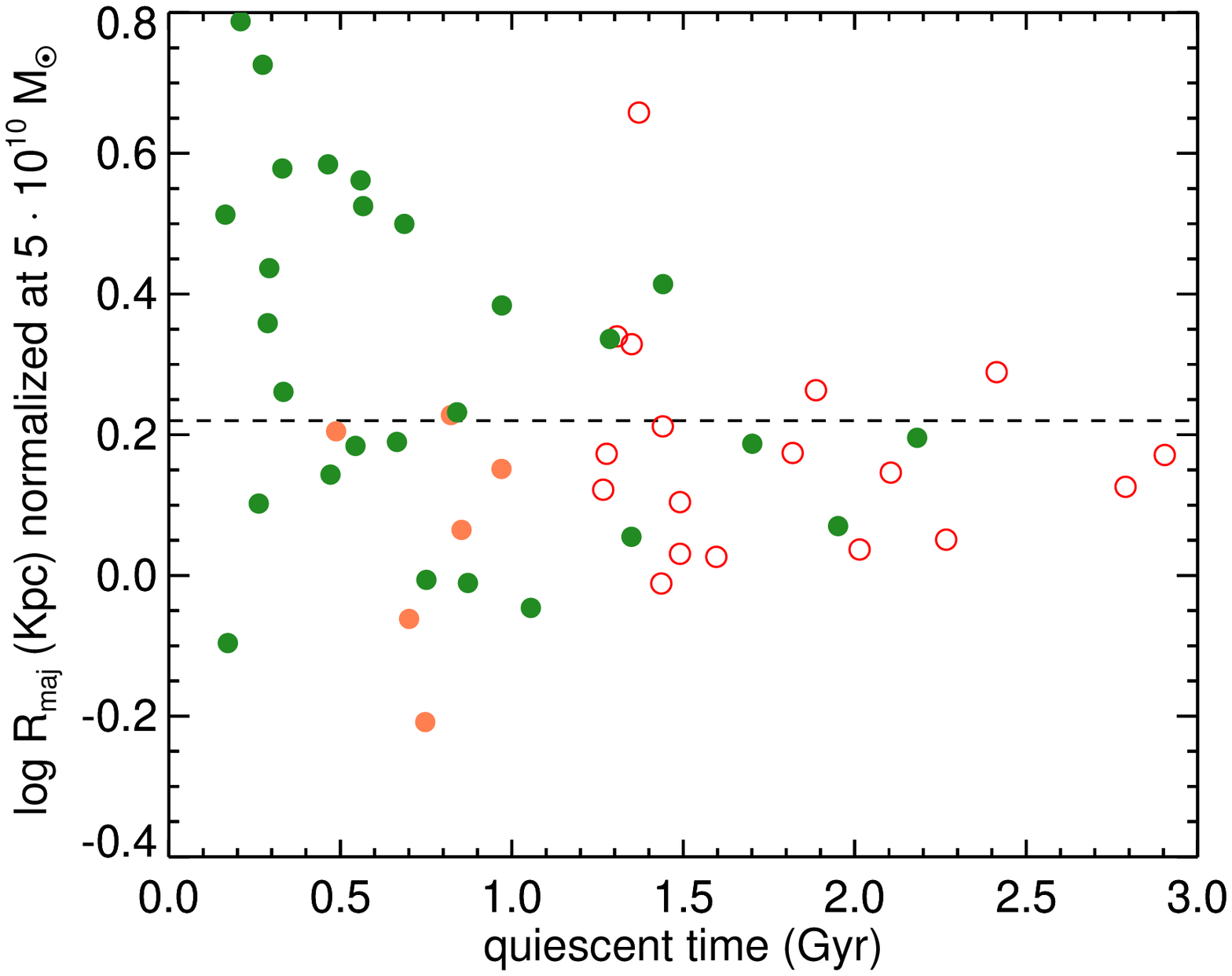}
\caption{Left: Stellar mass-size relation for quiescent galaxies in the LRIS sample. The color indicates the quiescent time $t_q$, which is the interval since the object became quiescent. The dashed line represents the relation derived from the 3D-HST sample at $z \sim 1.25$ \citep{vanderwel14}, and the solid line is the relation obtained for our sample assuming the same slope. Right: Size trends with quiescent time for both red sequence and green valley galaxies. The ordinate represents the size normalized to a fixed stellar mass of $5\cdot10^{10}\Msun$ using the mass-size relation shown in the left panel. The dashed line represents the median mass-normalized size: galaxies above this line lie above the mass-size relation. The points are color coded according to their stellar population properties as discussed in Section \ref{sec:redsequence_diversity}: old red sequence (red), post-starburst galaxies (orange), and green valley (green).}
\label{fig:mass-size}
\end{figure*}

We are now in a position to understand the rate at which the population of quiescent population is being enriched by recent arrivals. For each past time step we count the number of quiescent objects defined as above (numbers shown in black in each panel). Out of 51 quiescent galaxies at the epoch of observation, only 12 have been quiescent for more than 1.5 Gyr, thus the population grew by roughly a factor of four in a short period. Given we have shown that our sample is representative (Appendix \ref{sec:completeness}), we can thus compare the rate at which the quiescent population is growing from our simulated evolution to the results of photometrically-based stellar mass function studies, which are approximately volume-limited.

\citet{muzzin13} derive the stellar mass function for quiescent and star-forming galaxies over $0 < z < 4$ using a $UVJ$ color selection. This definition of the quiescent sample is in excellent agreement with the specific star formation rate threshold that we adopt, as we already discussed and as also evident from Figure \ref{fig:redseq_evo}. Using the Schechter function fits from \citet{muzzin13} we integrate over stellar masses larger than our adopted limit, $10^{10.7}\Msun$, to yield $\Phi(t)$, the number density of massive quiescent galaxies per unit comoving Mpc$^3$, as a function of cosmic time (Figure \ref{fig:redseq_growth_compare}). This must be compared to the number evolution inferred from the star formation histories of our spectroscopic sample, shown in red, up to the median epoch of observation at $z=1.25$. As we cannot rigorously calculate the cosmic volume probed by our spectroscopic observations, there is an unknown vertical offset in Figure \ref{fig:redseq_growth_compare}. Thus we should compare only the {\it rate of increase} in the quiescent population, which is in remarkable agreement with the mass function results. To estimate the uncertainty, we recalculate the number evolution many times, using slightly different star formation histories extracted from the posterior distribution of each galaxy, and plot the 68\% confidence region in light red. From our analysis we obtain a number density growth rate from $z=1.75$ to $z=1.25$ of $0.39 \pm 0.03$ dex, which compares favorably to $0.34 \pm 0.11$ derived from the stellar mass function study. This growth rate is not particularly sensitive to the selection of the quiescent population: shifting the $UVJ$ selection box of $\pm0.1$ mag changes the rate derived from the Muzzin et al. data by less than 0.08 dex. We note that our comparison neglects the effect of galaxy mergers, which can increase the stellar mass of quiescent galaxies that are just below the mass threshold, thus causing a growth in the number of massive quiescent galaxies that is not due to quenching. However, at this redshift the merger rate is much smaller than the quenching rate \citep[e.g.,][]{newman12}, and this effect can be neglected. 

The agreement between the number growth of the quiescent population that we reconstruct and the one directly observed as a function of redshift suggests our best-fit star formation histories are a reasonable description of the actual evolution of quiescent galaxies.

% ***************************************************************************************************
%						ANALYSIS: SIZE-AGE
% ***************************************************************************************************

\section{Size Evolution on the Red Sequence}
\label{sec:size}

We have used our technique to reconstruct the development of the quiescent population over a period of 1.5 Gyr prior to the median epoch of observation. This corresponds roughly to the redshift range $1.25 < z < 2$, where the size growth rate is particularly rapid. We are thus now in a position to directly estimate how recently-quenched galaxies that arrive on the red sequence during this time interval affect the size growth. In measuring physical sizes $R_\mathrm{maj}$ (effective radii measured along the major axis, listed in Table \ref{tab:sample}) for the LRIS sample, we use the methods described in detail in Paper I.

\subsection{The Size-Age Relation}
\label{sec:size-age}

Figure \ref{fig:mass-size} (left panel) shows the stellar mass-size relation for the quiescent galaxies in our sample. For convenience we compare this to the relation found at $z\sim1.25$ by \citet{vanderwel14} using the 3D-HST data (dashed line) as this survey also selected quiescent galaxies via their $UVJ$ colors. Although there is significant scatter, assuming the same slope we find the normalization for our sample differs from that for 3D-HST by only 0.01 dex (as shown by the solid line). The data points are color-coded according to their {\it quiescent time} $t_q$, defined as the time interval since the object's specific star formation rate fell below \tenminusten, following the discussion in Section 4.2. The value of $t_q$ is uniquely determined by age and $\tau$, as shown in Figure \ref{fig:tau-age} (red lines). Figure \ref{fig:mass-size} shows that galaxies which have been quiescent the longest, i.e. with the largest $t_q$, are physically more compact.

In the right panel of Figure \ref{fig:mass-size} we plot the deviation of galaxies from the mean mass-size relation as a function of their quiescent time. The deviation is simply the the vertical distance of each data point to the dashed line in Figure \ref{fig:mass-size}, normalized to the mean size at $5 \cdot 10^{10} \Msun$. In the right panel, points above the dashed line indicate galaxies which lie above the mass-size relation. Here we color code the galaxies according to whether they lie in the green valley, in the red sequence, or in the post-starburst region. This figure shows two important points. First, as we already saw in the left panel, older galaxies tend to be smaller, and vice-versa. Second, we now see that among young galaxies, the ones on the green valley are significantly larger than the post-starburst systems. In fact, the young and old halves of the red sequence have quite similar size distributions.

\begin{figure*}[tbp]
\centering
\includegraphics[width=0.7\textwidth]{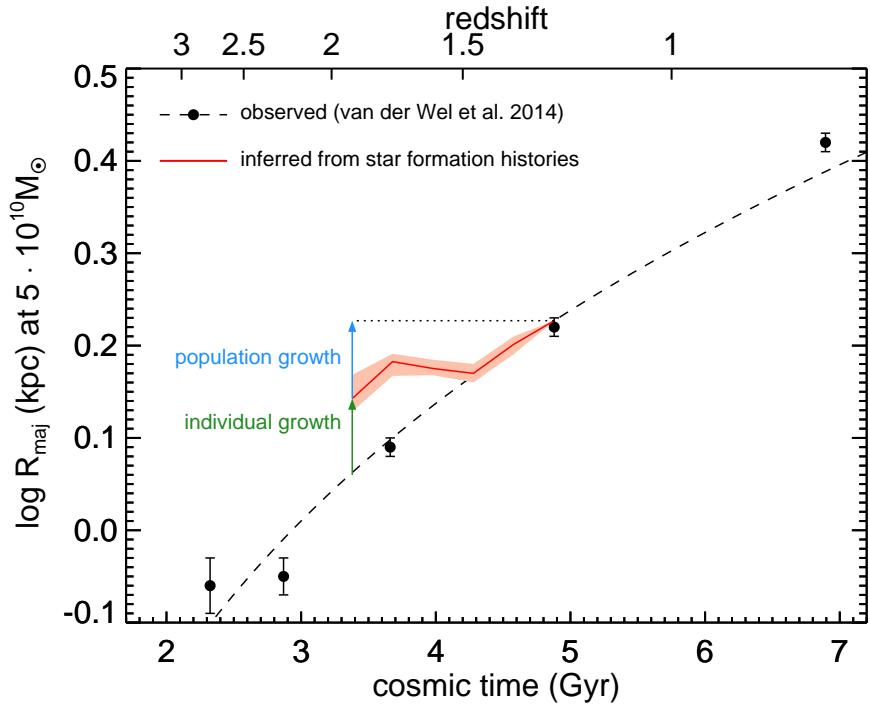}
\caption{Growth measured in terms of the normalization of the stellar mass-size relation for red sequence galaxies (parameterized as the average size at $\Mstar = 5 \cdot 10^{10} \Msun$), as a function of cosmic time. Black points represent the observations of \citet{vanderwel14} and the dashed line is their fit to the data. The red line is the evolution inferred by measuring the sizes of the galaxies in our sample that were quiescent at a given cosmic time. The effect of taking different star formation histories that are still consistent with the observations is shown by the shaded area. Our method is sensitive only to the growth due to the change in the composition of the quiescent population (blue arrow). The difference with the observed overall size evolution, then, must be due to the growth of individual galaxies (green arrow).}
\label{fig:redseq_sizegrowth_compare}
\end{figure*}

\subsection{The Contribution of Progenitor Bias to the Size Growth}

The overall goal of this paper is to use our reconstructed history of the red sequence to separate two modes of size growth in the redshift interval $1.25 < z < 2$. We will use the term \emph{individual size growth} to indicate a genuine increase in size for galaxies that have been on the red sequence throughout this period. \emph{Population size growth}, instead, refers to the apparent growth in size of red sequence galaxies arising from more recent arrivals which were larger prior to their quenching; this is the contribution from progenitor bias. As we have seen in the previous section, the oldest quiescent galaxies are typically the most compact and so, given we can reconstruct the rate of arrival of newly-quenched systems following our analysis in Section 4.2, we are ready to quantify the two modes of size growth. 

In Figure \ref{fig:redseq_sizegrowth_compare} we illustrate the size evolution via a red line, that we obtain in the same way as for the red line in Figure \ref{fig:redseq_growth_compare}, but measuring at each time step the average size (as opposed to just counting the number) of the quiescent galaxies. Again, the shaded area is obtained by varying the star formation histories according to the posterior distributions. The black points in the figure represent the evolution with redshift in the normalization of the mass-size relation from \citet{vanderwel14}, and the dashed line is a fit to the points. Since we are principally interested in the growth rate, we normalize the red line so it matches the \citet{vanderwel14} fit at $z=1.25$. This required shift is negligible as we already showed that our mass-size relation is in close absolute agreement with that of \citet{vanderwel14}.

Figure \ref{fig:redseq_sizegrowth_compare} shows the principal result of our study: size evolution due to the arrival of larger, newly-quenched galaxies - i.e. `population growth' - is insufficient to explain the observations. The size evolution of quiescent galaxies directly observed is $0.167 \pm 0.014$ dex over a 1.5 Gyr period, which is larger than that obtained above by measuring the sizes of the oldest galaxies at $z\sim1.25$, $0.084 \pm 0.020$ dex over the same period. The remainder ($0.083 \pm 0.024$ dex) must be due to individual size growth in long-standing quiescent objects. We show the relative contributions of individual and population size growth in Figure \ref{fig:redseq_sizegrowth_compare} with, respectively, blue and green arrows. In linear units, each process causes a relative size increase, at fixed mass, of 21\% over 1.5 Gyr. A more direct way to view this is to see that even the oldest, smallest objects at $z < 1.5$ are larger than the average quiescent galaxies observed at $z > 1.5$, a point first made by \citet{newman12}, which estimated the minimum individual growth by measuring the size increase of the smallest quiescent objects, obtaining a value in agreement with ours ($0.096 \pm 0.018$ dex over 1.5 Gyr). The only possible explanation for this difference is that physical growth of individual quiescent galaxies has occurred.

This result is very robust in terms of size measurements, which are accurate to the 10\% level for both our sample and the 3D-HST reference sample \citep{newman12,vanderwel12}. Due to the high quality of our spectroscopic data, this result is also robust against random errors in the age estimates, as shown in Figure \ref{fig:redseq_sizegrowth_compare}. These do not include systematic effects, due, e.g., to the assumption of simple declining star formation histories, which do not include the effect of secondary bursts. However, the agreement between our reconstructed number evolution of the red sequence with the evolution directly observed by \citet[][shown in Figure \ref{fig:redseq_growth_compare}]{muzzin13} strongly suggests that our ages are not significantly biased. Regarding the size evolution we also made the implicit assumption that the \emph{observed} size of a galaxy does not change during the quenching process. The size, however, might decrease because of disk instability (that causes a change in the mass distribution) or because of the removal of dust (which would cause a change in the light distribution). In both cases the effect of newly quenched galaxies on the mean mass-size relation would be smaller than what assumed in our analysis, and therefore our measurement of the progenitor bias would be an upper limit.

% ***************************************************************************************************
%						SUMMARY AND DISCUSSION
% ***************************************************************************************************

\section{Summary and Discussion}
\label{sec:discussion}

Taking advantage of deep LRIS spectra, together with associated imaging and broad-band photometry, we have investigated the stellar population parameters of an unbiased sample of quiescent galaxies within the redshift range $1 < z < 1.6$. By reconstructing their star formation histories, we were able to reproduce the evolution in number density of quiescent galaxies measured independently in deep photometric surveys. We measured the relation between size and mass, and found that older galaxies are significantly smaller. We then reconstructed the evolution of the mean size in the 1.5 Gyr prior to the time of observation. Comparing this to the mean sizes measured at different redshifts from the HST CANDELS survey, we found that the oldest galaxies in our sample must have been growing in size since $z\sim2$.

Our result is in agreement with the conclusions of dynamical studies undertaken at higher redshift. In \citet{belli14mosfire}, we measured velocity dispersions for a small sample of quiescent galaxies at $2 < z < 2.5$, and by comparing their sizes and masses to those of local galaxies with same velocity dispersion, we concluded that physical growth occurred. It is noteworthy that the physical growth of quiescent galaxies over the period corresponding to $1 < z < 2$, first suggested by number density arguments \citep[e.g.,][]{bezanson09,newman12}, has now been confirmed by two independent techniques and datasets.

As the apparent growth over $1.25 < z < 2$ can now be dissected into a near-equal combination of genuine (physical) growth and that arising from recently-quenched arrivals (progenitor bias), the question arises as to the mechanism by which the older quiescent galaxies are growing. In \citetalias{belli14lris} we showed that minor mergers are likely to be the primary mechanism for the size growth over $0 < z < 1.5$ \citep[see also][]{vandokkum10, nipoti12, posti14}. However, at $z\sim2$ spectroscopic observations suggest that the growth in mass and size is inconsistent with arising exclusively via minor mergers \citep{belli14mosfire}. Moreover, the merger rate inferred from \HST\ imaging \citep{newman12} appears to be insufficient to account for the physical growth even after accounting for progenitor bias. Hopefully improved estimates of the minor merger rate together with larger spectroscopic samples beyond $z\sim2$ will enable us to address this important remaining question in the evolution of compact quiescent galaxies.

Our result was made possible by the high quality of the spectroscopic data, which allowed us to derive accurate stellar population parameters. An earlier attempt to measure the relation between size and age at $1 < z  < 2$ used the $UVJ$ colors as proxy for age. By splitting the red sequence into blue and red halves, \citet{whitaker12} did not detect any difference in size. As we showed in Section \ref{sec:redsequence}, the post-starburst objects that populate the blue side of the red sequence do, in fact, show similar sizes to the oldest galaxies. The main contribution to the size growth of the population comes instead from galaxies in the green valley.

\subsection{Two Pathways to Quenching?}

One of the unexpected findings of this study was the distinction between green valley galaxies and post-starburst systems, both of which lie within the quiescent population defined in Section \ref{sec:stellarpop}. Given the different levels of star formation rate for these two populations, one might conclude that green valley and post-starburst phase represent successive stages in the overall evolution from the blue cloud to the red sequence. This is clearly not the case. All the post-starburst galaxies have ages around 1 Gyr, and very small values of $\tau$, therefore their quiescent times are also around or slightly below 1 Gyr (see Figure \ref{fig:tau-age}). However, green valley galaxies have ages between 1 and 4 Gyr, and quiescent times that span the entire range between 0 and 4 Gyr. Our data are inconsistent with a simple picture in which quenched galaxies first cross the green valley before moving through a post-starburst phase and arriving on the red sequence. The more likely explanation  is one in which the green valley and the post-starburst phase represent two independent evolutionary paths. The main difference is the quenching timescale: the low values of $\tau$ for post-starburst galaxies correspond to a fast quenching, whereas for the green valley galaxies, $\tau$ is comparable to the age, resulting in slowly declining star formation rates. This difference in timescales results in different levels of star formation rates for galaxies of identical ages.

Interestingly, this picture is consistent with the studies of \citet{patel13} and \citet{marchesini14} which follow the evolution of a galaxy population by matching number densities at different redshifts. These authors find that at high redshift the progenitors of local massive quiescent galaxies are located both on the blue end of the red sequence and on the green valley. More importantly, the progenitors on the red sequence move toward the red end with cosmic time, while at the same time the green valley remains significantly populated. This implies that the post-starburst phase is not just the endpoint of the evolution of green valley galaxies, but constitutes an independent path, which in the case of ultra-massive galaxies ends by $z \sim 1.5$ (see, e.g., Figure 2 of \citealt{marchesini14}).

The star formation histories are not the only properties that are distinct across the two quiescent sub-populations. Green valley systems are typically large and dusty, while post-starburst galaxies have little dust and smaller sizes. Although the best-fit dust extinction can be degenerate with stellar population ages, the sizes are clearly independently measured. Furthermore, using independent mid-IR emission as a proxy for dust extinction does not significantly change our results, thus confirming the robustness of our conclusions (see Figure \ref{fig:uvj}).

The possibility of two quenching channels with different timescales has also been proposed at low redshift by \citeauthor{schawinski14} \citep[2014, see also][]{yesuf14}, who suggest that major mergers produce a fast quenching and a morphological transformation, while the slow quenching might be caused by some process, such as AGN feedback, that interrupts gas accretion. On the theoretical side, a number of simulations are consistent with quenching being caused by two essentially unrelated physical processes \citep[e.g.,][]{woo14,wellons14}.

Potential progenitors of compact quiescent galaxies have been identified by \citet{barro13}, which selected a sample of compact star-forming galaxies at $z > 2$.  Among these galaxies, the ones near the blue end of the $UVJ$ red sequence tend to be small and dust-free \citep{barro14}. These objects are likely to be the immediate progenitors of the post-starburst systems that we identified at $z < 1.5$. Dusty star-forming objects, such as sub-mm galaxies, might on the other hand be the progenitors of the galaxies on the green valley \citep{toft14,nelson14}. However, further studies of transitional galaxies, including more detailed analysis of their star formation histories and morphologies, are needed in order to understand the physical processes responsible for galaxy quenching. \\

% ***************************************************************************************************

We acknowledge Danilo Marchesini for useful discussions. The authors recognize and acknowledge the very significant cultural role and reverence that the summit of Mauna Kea has always had within the indigenous Hawaiian community. We are most fortunate to have the opportunity to conduct observations from this mountain.

% ***************************************************************************************************
%						APPENDIX: COMPLETENESS OF THE SAMPLE
% ***************************************************************************************************

\appendix
\section{The Spectroscopic Sample is Unbiased}
\label{sec:completeness}

Spectroscopic samples are typically biased, because of the combined effects of target selection and the need to identify spectral features. It is therefore critical to assess whether our sample is biased. For this purpose, it is necessary to use a larger catalog that can be considered complete down to masses below $\sim 10^{10.7}\Msun$. For this task we use the public catalog from the 3D-HST survey \citep{brammer12, skelton14}, that presents two important advantages. Firstly, it was obtained in the same CANDELS fields in which the majority of our targets lie, allowing a more direct comparison; secondly, the 3D-HST team adopts the $UVJ$ plane for dividing galaxies into quiescent and star-forming, and this ensures consistency in the definition of the samples.

The 3D-HST catalog contains, among other properties, photometric redshift, stellar mass, and rest-frame colors for every object. We selected all the objects with $1 < z < 1.6$ and $\Mstar > 10^{10.7}\Msun$, and call this the reference sample. We also identify 58 of our 62 objects in the 3D-HST catalog, by matching the coordinates. Rather than comparing the properties that we derived for our objects with those published for the reference sample, we carry out a self-consistent comparison by using only the properties from the 3D-HST catalog.

\begin{figure*}[t!]
    \centering
    \begin{subfigure}
        \centering
        \includegraphics[width=0.45\textwidth]{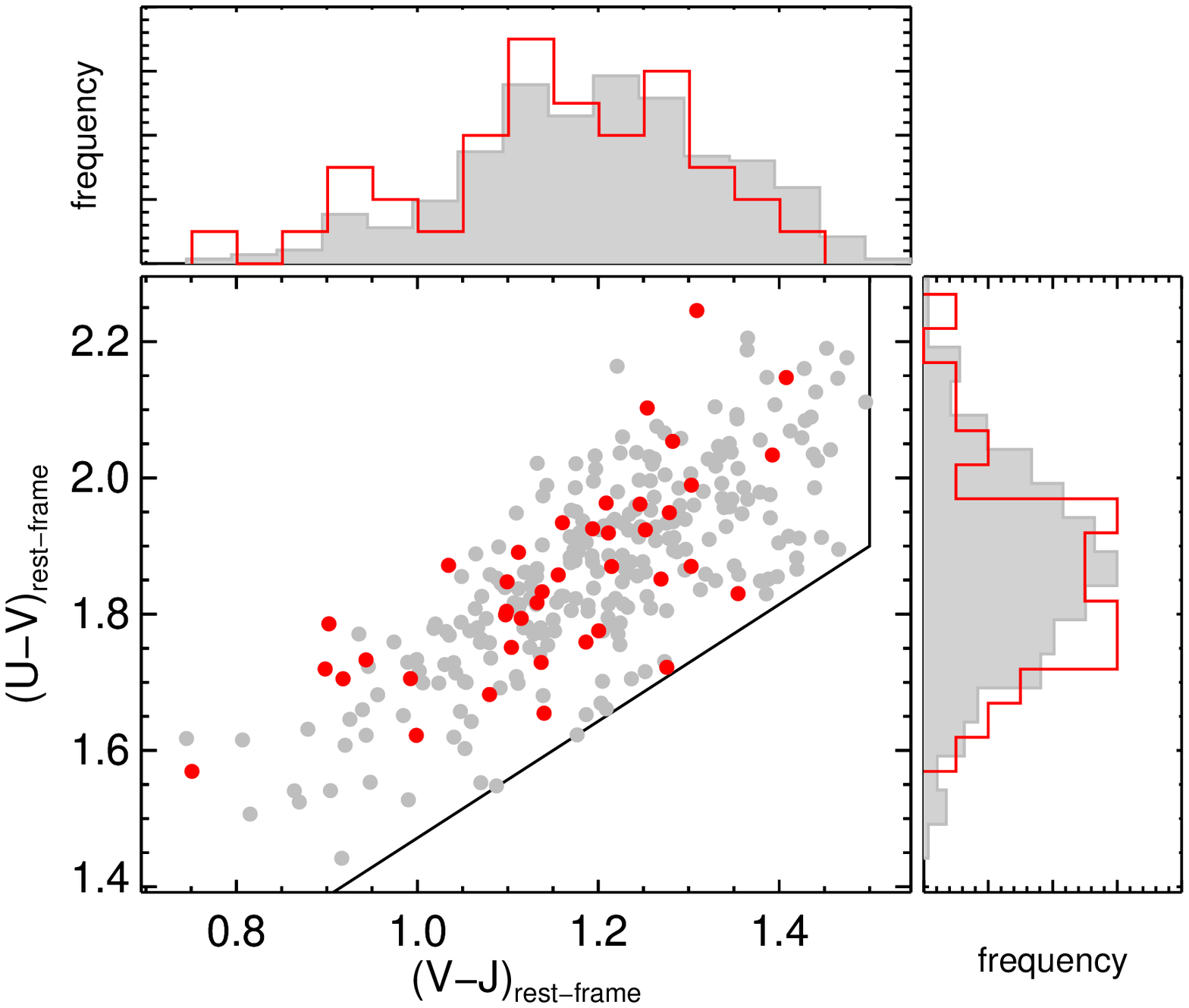}
    \end{subfigure}%
    \begin{subfigure}
        \centering
        \includegraphics[width=0.45\textwidth]{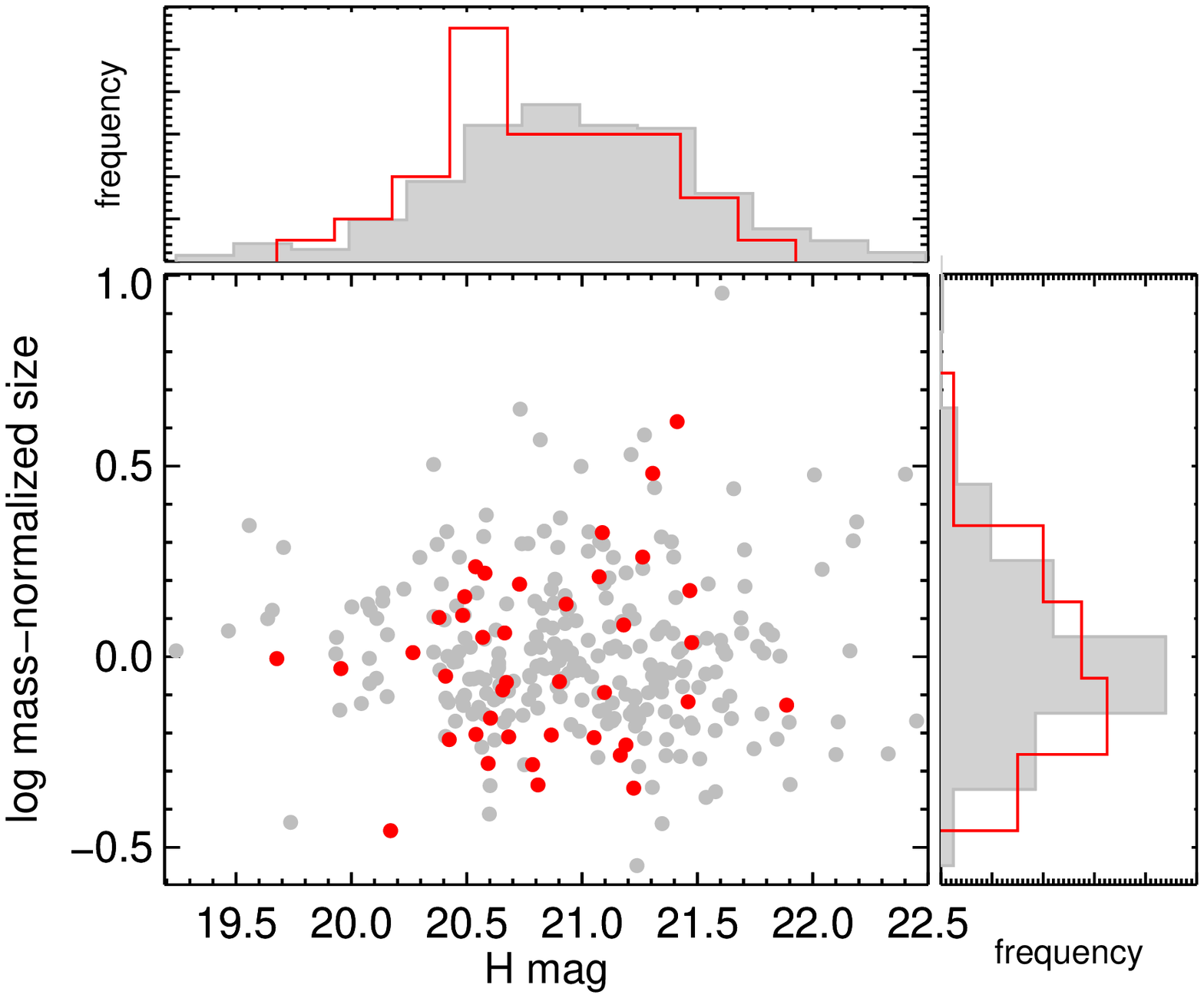}
    \end{subfigure}
    \caption{Left: Comparison of the distribution on the $UVJ$ diagram for our sample (red) and for the the 3D-HST reference sample, defined by $\log \Mstar/\Msun > 10.7$ and $1 < z < 1.6$ (gray). Only galaxies in the quiescent selection box are shown. The top and right panels show the histograms of the rest-frame colors for the two samples. Right: Comparison of the magnitude and mass-normalized distributions. In both panels, the properties of our sample are taken from the 3D-HST catalog, to ensure a consistent comparison.}
\label{fig:samplebias}
\end{figure*}

The left panel of Figure \ref{fig:samplebias} shows the distribution of our sample (red points) and the reference sample (gray points) in the $UVJ$ diagram. Only the objects in the quiescent selection box are shown. The two histograms compare the rest-frame $U-V$ and $V-J$ colors for our sample and for the reference population. The two samples are remarkably similar, and a K-S test confirms that the two distributions are formally consistent with each other, in both $V-J$ ($p=0.43$) and $U-V$ ($p=0.63$).

We note that when we use the rest-frame colors derived from our best-fit models we obtain slightly different results. Comparing the colors calculated by us to the ones calculated by the 3D-HST team \emph{for the same objects} in our sample, we find a mean shift $\Delta(V-J) = 0.12$ and $\Delta(U-V) = 0.03$. This discrepancy is probably caused by a difference in the templates used: we calculate the colors by integrating our best-fit template, while the 3D-HST colors are obtained from the {\tt EAZY} templates \citep{brammer08}, which include emission lines. As a consequence, the sample shown in Figure \ref{fig:samplebias} is slightly different from the sample used in the rest of the present paper, as the slightly different rest-frame colors can cause some objects to fall inside or outside the selection box. We note that the star-forming galaxies are the ones most affected by this issue, while the objects on the red sequence show the smallest discrepancy.

In the right panel of Figure \ref{fig:samplebias} we compare the distribution of our sample in $H$ magnitude and mass-normalized size with the reference sample. Again, we can see that our spectroscopic sample is unbiased compared to the parent population, as is confirmed by the K-S test ($p=0.29$ for the $H$ distibutions and $p=0.19$ for the size distributions).

We conclude, therefore, that our sample of quiescent galaxies is unbiased, and represents well the underlying galaxy population.

\begin{deluxetable*}{llccccccc}
\tabletypesize{\footnotesize}
\tablewidth{0pc}
\tablecaption{Stellar Population Properties of the Sample of Quiescent Galaxies \label{tab:sample}}
\tablehead{
\colhead{Object ID} & \colhead{$z$} & \colhead{log sSFR} & \colhead{$\log \Mstar$} & \colhead{log Age} & \colhead{$\log \tau$} & \colhead{$A_V$} & \colhead{$Z/0.02$} & $R_\mathrm{maj}$
\\
 & & (yr$^{-1}$) & (\Msun) & (yr) & (yr) & & & (kpc) 
} 
\startdata
19826 & 1.008 & $ -10.55 \pm 0.06 $ & $ 11.07 \pm 0.04 $ & $ 9.63 \pm 0.06 $ & $ 9.01 \pm 0.06 $ & $ 0.13 \pm 0.04 $ & $ 0.44 \pm 0.04 $ & $ 2.2 $  \\
51106 & 1.013 & $ -11.18 \pm 0.09 $ & $ 11.29 \pm 0.03 $ & $ 9.37 \pm 0.05 $ & $ 8.55 \pm 0.06 $ & $ 0.75 \pm 0.08 $ & $ 0.89 \pm 0.17 $ & $ 6.8 $  \\
28739 & 1.029 & $ < -12 $ & $ 11.03 \pm 0.03 $ & $ 9.54 \pm 0.04 $ & $ 8.15 \pm 0.55 $ & $ 0.11 \pm 0.05 $ & $ 0.66 \pm 0.10 $ & $ 2.4 $  \\
21741 & 1.055 & $ < -12 $ & $ 10.92 \pm 0.03 $ & $ 9.24 \pm 0.05 $ & $ 7.85 \pm 0.42 $ & $ 0.41 \pm 0.07 $ & $ 0.74 \pm 0.17 $ & $ 3.1 $  \\
49418 & 1.061 & $ < -12 $ & $ 11.37 \pm 0.04 $ & $ 9.48 \pm 0.07 $ & $ 8.47 \pm 0.46 $ & $ 0.17 \pm 0.09 $ & $ 0.79 \pm 0.17 $ & $ 4.8 $  \\
51081 & 1.062 & $ -10.37 \pm 0.10 $ & $ 10.96 \pm 0.04 $ & $ 9.26 \pm 0.06 $ & $ 8.60 \pm 0.07 $ & $ 0.71 \pm 0.09 $ & $ 0.91 \pm 0.18 $ & $ 6.0 $  \\
31377 & 1.085 & $ -10.43 \pm 0.13 $ & $ 10.70 \pm 0.02 $ & $ 9.06 \pm 0.02 $ & $ 8.34 \pm 0.04 $ & $ 1.42 \pm 0.09 $ & $ 0.61 \pm 0.15 $ & $ 6.2 $  \\
13393 & 1.097 & $ -10.52 \pm 0.05 $ & $ 11.15 \pm 0.03 $ & $ 9.34 \pm 0.05 $ & $ 8.66 \pm 0.05 $ & $ 0.56 \pm 0.06 $ & $ 0.67 \pm 0.17 $ & $ 8.0 $  \\
16343 & 1.098 & $ -11.84 \pm 0.19 $ & $ 11.01 \pm 0.01 $ & $ 9.02 \pm 0.01 $ & $ 8.06 \pm 0.09 $ & $ 0.32 \pm 0.03 $ & $ 1.02 \pm 0.03 $ & $ 2.4 $  \\
28656 & 1.101 & $ < -12 $ & $ 11.19 \pm 0.03 $ & $ 9.55 \pm 0.04 $ & $ 8.59 \pm 0.40 $ & $ 0.06 \pm 0.04 $ & $ 0.79 \pm 0.09 $ & $ 3.3 $  \\
32591 & 1.110 & $ < -12 $ & $ 11.36 \pm 0.02 $ & $ 9.51 \pm 0.03 $ & $ 7.71 \pm 0.49 $ & $ 0.04 \pm 0.05 $ & $ 1.01 \pm 0.05 $ & $ 4.7 $  \\
21715 & 1.113 & $ -10.85 \pm 0.07 $ & $ 10.92 \pm 0.03 $ & $ 9.30 \pm 0.03 $ & $ 8.53 \pm 0.04 $ & $ 0.44 \pm 0.05 $ & $ 1.06 \pm 0.12 $ & $ 2.3 $  \\
21657 & 1.125 & $ -11.38 \pm 0.09 $ & $ 11.13 \pm 0.04 $ & $ 9.60 \pm 0.05 $ & $ 8.78 \pm 0.06 $ & $ 0.23 \pm 0.07 $ & $ 1.14 \pm 0.17 $ & $ 2.5 $  \\
12988 & 1.144 & $ -10.33 \pm 0.05 $ & $ 10.97 \pm 0.02 $ & $ 9.29 \pm 0.03 $ & $ 8.65 \pm 0.04 $ & $ 0.46 \pm 0.05 $ & $ 0.84 \pm 0.16 $ & $ 3.0 $  \\
1672 & 1.147 & $ -10.44 \pm 0.07 $ & $ 11.05 \pm 0.03 $ & $ 9.14 \pm 0.04 $ & $ 8.43 \pm 0.05 $ & $ 0.95 \pm 0.06 $ & $ 1.18 \pm 0.17 $ & $ 9.8 $  \\
21870 & 1.179 & $ < -12 $ & $ 11.11 \pm 0.02 $ & $ 9.35 \pm 0.03 $ & $ 7.83 \pm 0.43 $ & $ 0.26 \pm 0.05 $ & $ 0.53 \pm 0.08 $ & $ 3.8 $  \\
1241357 & 1.188 & $ < -12 $ & $ 10.90 \pm 0.02 $ & $ 9.50 \pm 0.03 $ & $ 8.29 \pm 0.46 $ & $ 0.03 \pm 0.03 $ & $ 0.89 \pm 0.11 $ & $ 1.6 $  \\
41327 & 1.192 & $ < -12 $ & $ 10.82 \pm 0.03 $ & $ 8.98 \pm 0.04 $ & $ 7.96 \pm 0.32 $ & $ 0.46 \pm 0.10 $ & $ 0.84 \pm 0.28 $ & $ 2.0 $  \\
33887 & 1.193 & $ -10.66 \pm 0.10 $ & $ 10.88 \pm 0.04 $ & $ 9.36 \pm 0.06 $ & $ 8.65 \pm 0.07 $ & $ 0.31 \pm 0.09 $ & $ 1.05 \pm 0.23 $ & $ 4.4 $  \\
45759 & 1.196 & $ -10.37 \pm 0.12 $ & $ 10.92 \pm 0.05 $ & $ 9.35 \pm 0.08 $ & $ 8.70 \pm 0.09 $ & $ 0.53 \pm 0.12 $ & $ 0.96 \pm 0.20 $ & $ 5.6 $  \\
3346 & 1.217 & $ -10.75 \pm 0.08 $ & $ 10.80 \pm 0.02 $ & $ 9.04 \pm 0.04 $ & $ 8.25 \pm 0.04 $ & $ 0.51 \pm 0.09 $ & $ 1.05 \pm 0.24 $ & $ 3.3 $  \\
3867 & 1.223 & $ -11.59 \pm 0.11 $ & $ 10.82 \pm 0.02 $ & $ 9.47 \pm 0.04 $ & $ 8.59 \pm 0.09 $ & $ 0.07 \pm 0.04 $ & $ 0.62 \pm 0.12 $ & $ 3.2 $  \\
34609 & 1.241 & $ < -12 $ & $ 11.02 \pm 0.02 $ & $ 9.20 \pm 0.04 $ & $ 7.53 \pm 0.37 $ & $ 0.80 \pm 0.06 $ & $ 0.81 \pm 0.15 $ & $ 7.9 $  \\
21750 & 1.242 & $ -11.51 \pm 0.10 $ & $ 11.10 \pm 0.03 $ & $ 9.27 \pm 0.05 $ & $ 8.38 \pm 0.05 $ & $ 0.31 \pm 0.06 $ & $ 0.96 \pm 0.22 $ & $ 3.4 $  \\
7662 & 1.244 & $ < -12 $ & $ 10.95 \pm 0.00 $ & $ 9.30 \pm 0.00 $ & $ 7.85 \pm 0.00 $ & $ 0.21 \pm 0.00 $ & $ 0.98 \pm 0.00 $ & $ 1.7 $  \\
18249 & 1.252 & $ -11.18 \pm 0.12 $ & $ 10.81 \pm 0.04 $ & $ 9.17 \pm 0.05 $ & $ 8.32 \pm 0.06 $ & $ 0.49 \pm 0.06 $ & $ 1.24 \pm 0.19 $ & $ 4.0 $  \\
7310 & 1.255 & $ < -12 $ & $ 11.13 \pm 0.02 $ & $ 9.21 \pm 0.03 $ & $ 7.76 \pm 0.40 $ & $ 0.60 \pm 0.05 $ & $ 0.54 \pm 0.13 $ & $ 4.7 $  \\
13073 & 1.258 & $ -11.50 \pm 0.09 $ & $ 11.01 \pm 0.02 $ & $ 9.23 \pm 0.03 $ & $ 8.34 \pm 0.03 $ & $ 0.18 \pm 0.04 $ & $ 0.90 \pm 0.15 $ & $ 1.7 $  \\
30822 & 1.259 & $ < -12 $ & $ 10.99 \pm 0.04 $ & $ 9.27 \pm 0.08 $ & $ 8.07 \pm 0.45 $ & $ 0.48 \pm 0.10 $ & $ 0.97 \pm 0.43 $ & $ 2.2 $  \\
1244914 & 1.261 & $ -11.22 \pm 0.05 $ & $ 11.24 \pm 0.02 $ & $ 9.47 \pm 0.03 $ & $ 8.66 \pm 0.04 $ & $ 0.19 \pm 0.05 $ & $ 0.94 \pm 0.19 $ & $ 5.6 $  \\
32915 & 1.261 & $ -11.01 \pm 0.05 $ & $ 10.98 \pm 0.02 $ & $ 9.43 \pm 0.03 $ & $ 8.65 \pm 0.04 $ & $ 0.17 \pm 0.05 $ & $ 0.54 \pm 0.11 $ & $ 1.5 $  \\
22760 & 1.262 & $ < -12 $ & $ 10.90 \pm 0.01 $ & $ 9.36 \pm 0.02 $ & $ 7.66 \pm 0.37 $ & $ 0.33 \pm 0.06 $ & $ 0.54 \pm 0.09 $ & $ 1.5 $  \\
22780 & 1.264 & $ -10.77 \pm 0.10 $ & $ 10.77 \pm 0.01 $ & $ 9.03 \pm 0.02 $ & $ 8.23 \pm 0.03 $ & $ 0.62 \pm 0.06 $ & $ 0.84 \pm 0.14 $ & $ 2.6 $  \\
2341 & 1.266 & $ < -12 $ & $ 10.82 \pm 0.02 $ & $ 9.01 \pm 0.01 $ & $ 7.39 \pm 0.26 $ & $ 0.47 \pm 0.06 $ & $ 1.08 \pm 0.12 $ & $ 1.4 $  \\
29059 & 1.278 & $ -10.62 \pm 0.09 $ & $ 10.91 \pm 0.02 $ & $ 9.06 \pm 0.04 $ & $ 8.29 \pm 0.05 $ & $ 0.44 \pm 0.08 $ & $ 1.10 \pm 0.22 $ & $ 1.8 $  \\
2823 & 1.316 & $ -11.12 \pm 0.12 $ & $ 11.26 \pm 0.04 $ & $ 9.58 \pm 0.05 $ & $ 8.81 \pm 0.07 $ & $ 0.42 \pm 0.09 $ & $ 0.86 \pm 0.17 $ & $ 4.1 $  \\
34879 & 1.322 & $ -11.64 \pm 0.10 $ & $ 11.45 \pm 0.03 $ & $ 9.61 \pm 0.04 $ & $ 8.75 \pm 0.04 $ & $ 0.35 \pm 0.06 $ & $ 0.61 \pm 0.10 $ & $ 5.8 $  \\
2337 & 1.327 & $ < -12 $ & $ 11.06 \pm 0.06 $ & $ 9.23 \pm 0.07 $ & $ 7.67 \pm 0.41 $ & $ 0.34 \pm 0.06 $ & $ 0.75 \pm 0.25 $ & $ 1.8 $  \\
14758 & 1.331 & $ < -12 $ & $ 10.72 \pm 0.02 $ & $ 8.95 \pm 0.02 $ & $ 7.50 \pm 0.27 $ & $ 0.53 \pm 0.05 $ & $ 0.73 \pm 0.14 $ & $ 0.9 $  \\
33786 & 1.352 & $ -10.21 \pm 0.11 $ & $ 10.83 \pm 0.03 $ & $ 9.16 \pm 0.06 $ & $ 8.50 \pm 0.07 $ & $ 0.70 \pm 0.10 $ & $ 0.94 \pm 0.21 $ & $ 4.1 $  \\
25374 & 1.397 & $ < -12 $ & $ 10.90 \pm 0.09 $ & $ 9.44 \pm 0.10 $ & $ 7.78 \pm 0.47 $ & $ 0.20 \pm 0.16 $ & $ 0.83 \pm 0.23 $ & $ 2.8 $  \\
19498 & 1.401 & $ < -12 $ & $ 10.78 \pm 0.04 $ & $ 9.25 \pm 0.07 $ & $ 7.68 \pm 0.42 $ & $ 0.19 \pm 0.07 $ & $ 0.79 \pm 0.25 $ & $ 1.2 $  \\
5835 & 1.405 & $ -10.35 \pm 0.07 $ & $ 10.93 \pm 0.02 $ & $ 9.42 \pm 0.05 $ & $ 8.80 \pm 0.05 $ & $ 0.35 \pm 0.07 $ & $ 0.58 \pm 0.15 $ & $ 2.3 $  \\
42109 & 1.406 & $ -11.29 \pm 0.08 $ & $ 10.79 \pm 0.03 $ & $ 9.32 \pm 0.05 $ & $ 8.47 \pm 0.05 $ & $ 0.08 \pm 0.06 $ & $ 0.62 \pm 0.14 $ & $ 1.2 $  \\
5020 & 1.415 & $ < -12 $ & $ 10.87 \pm 0.03 $ & $ 9.28 \pm 0.08 $ & $ 7.94 \pm 0.45 $ & $ 0.15 \pm 0.06 $ & $ 0.51 \pm 0.08 $ & $ 2.2 $  \\
4906 & 1.419 & $ < -12 $ & $ 11.13 \pm 0.07 $ & $ 9.05 \pm 0.08 $ & $ 7.39 \pm 0.27 $ & $ 0.57 \pm 0.08 $ & $ 1.04 \pm 0.11 $ & $ 3.0 $  \\
20275 & 1.442 & $ < -12 $ & $ 10.77 \pm 0.03 $ & $ 9.02 \pm 0.08 $ & $ 7.54 \pm 0.30 $ & $ 0.50 \pm 0.13 $ & $ 1.12 \pm 0.20 $ & $ 1.9 $  \\
40620 & 1.478 & $ < -12 $ & $ 11.17 \pm 0.03 $ & $ 9.24 \pm 0.06 $ & $ 7.60 \pm 0.34 $ & $ 0.18 \pm 0.07 $ & $ 0.67 \pm 0.16 $ & $ 2.9 $  \\
17468 & 1.529 & $ < -12 $ & $ 10.76 \pm 0.09 $ & $ 9.21 \pm 0.17 $ & $ 7.81 \pm 0.42 $ & $ 0.39 \pm 0.14 $ & $ 0.87 \pm 0.25 $ & $ 1.7 $  \\
34265 & 1.582 & $ < -12 $ & $ 11.36 \pm 0.01 $ & $ 8.96 \pm 0.02 $ & $ 7.41 \pm 0.24 $ & $ 0.52 \pm 0.04 $ & $ 0.95 \pm 0.09 $ & $ 2.0 $  \\
2653 & 1.598 & $ -10.60 \pm 0.31 $ & $ 10.94 \pm 0.02 $ & $ 8.91 \pm 0.04 $ & $ 8.12 \pm 0.09 $ & $ 0.68 \pm 0.11 $ & $ 0.83 \pm 0.15 $ & $ 1.2 $
\enddata
\end{deluxetable*}

% ***************************************************************************************************

\bibliography{sirio}

\begin{thebibliography}{74}
\expandafter\ifx\csname natexlab\endcsname\relax\def\natexlab#1{#1}\fi

\bibitem[{{Baldry} {et~al.}(2004){Baldry}, {Glazebrook}, {Brinkmann},
  {Ivezi{\'c}}, {Lupton}, {Nichol}, \& {Szalay}}]{baldry04}
{Baldry}, I.~K., {Glazebrook}, K., {Brinkmann}, J., {Ivezi{\'c}}, {\v Z}.,
  {Lupton}, R.~H., {Nichol}, R.~C., \& {Szalay}, A.~S. 2004, \apj, 600, 681

\bibitem[{{Barro} {et~al.}(2011){Barro}, {P{\'e}rez-Gonz{\'a}lez}, {Gallego},
  {Ashby}, {Kajisawa}, {Miyazaki}, {Villar}, {Yamada}, \& {Zamorano}}]{barro11}
{Barro}, G., {et~al.} 2011, \apjs, 193, 13

\bibitem[{{Barro} {et~al.}(2013){Barro}, {Faber}, {P{\'e}rez-Gonz{\'a}lez},
  {Koo}, {Williams}, {Kocevski}, {Trump}, {Mozena}, {McGrath}, {van der Wel},
  {Wuyts}, {Bell}, {Croton}, {Ceverino}, {Dekel}, {Ashby}, {Cheung},
  {Ferguson}, {Fontana}, {Fang}, {Giavalisco}, {Grogin}, {Guo}, {Hathi},
  {Hopkins}, {Huang}, {Koekemoer}, {Kartaltepe}, {Lee}, {Newman}, {Porter},
  {Primack}, {Ryan}, {Rosario}, {Somerville}, {Salvato}, \& {Hsu}}]{barro13}
---. 2013, \apj, 765, 104

\bibitem[{{Barro} {et~al.}(2014){Barro}, {Faber}, {P{\'e}rez-Gonz{\'a}lez},
  {Pacifici}, {Trump}, {Koo}, {Wuyts}, {Guo}, {Bell}, {Dekel}, {Porter},
  {Primack}, {Ferguson}, {Ashby}, {Caputi}, {Ceverino}, {Croton}, {Fazio},
  {Giavalisco}, {Hsu}, {Kocevski}, {Koekemoer}, {Kurczynski}, {Kollipara},
  {Lee}, {McIntosh}, {McGrath}, {Moody}, {Somerville}, {Papovich}, {Salvato},
  {Santini}, {Tal}, {van der Wel}, {Williams}, {Willner}, \&
  {Zolotov}}]{barro14}
---. 2014, \apj, 791, 52

\bibitem[{{Belli} {et~al.}(2014{\natexlab{a}}){Belli}, {Newman}, \&
  {Ellis}}]{belli14lris}
{Belli}, S., {Newman}, A.~B., \& {Ellis}, R.~S. 2014{\natexlab{a}}, \apj, 783,
  117

\bibitem[{{Belli} {et~al.}(2014{\natexlab{b}}){Belli}, {Newman}, {Ellis}, \&
  {Konidaris}}]{belli14mosfire}
{Belli}, S., {Newman}, A.~B., {Ellis}, R.~S., \& {Konidaris}, N.~P.
  2014{\natexlab{b}}, \apjl, 788, L29

\bibitem[{{Bezanson} {et~al.}(2009){Bezanson}, {van Dokkum}, {Tal},
  {Marchesini}, {Kriek}, {Franx}, \& {Coppi}}]{bezanson09}
{Bezanson}, R., {van Dokkum}, P.~G., {Tal}, T., {Marchesini}, D., {Kriek}, M.,
  {Franx}, M., \& {Coppi}, P. 2009, \apj, 697, 1290

\bibitem[{{Bielby} {et~al.}(2012){Bielby}, {Hudelot}, {McCracken}, {Ilbert},
  {Daddi}, {Le F{\`e}vre}, {Gonzalez-Perez}, {Kneib}, {Marmo}, {Mellier},
  {Salvato}, {Sanders}, \& {Willott}}]{bielby12}
{Bielby}, R., {et~al.} 2012, \aap, 545, A23

\bibitem[{{Blanton} {et~al.}(2003){Blanton}, {Hogg}, {Bahcall}, {Baldry},
  {Brinkmann}, {Csabai}, {Eisenstein}, {Fukugita}, {Gunn}, {Ivezi{\'c}},
  {Lamb}, {Lupton}, {Loveday}, {Munn}, {Nichol}, {Okamura}, {Schlegel},
  {Shimasaku}, {Strauss}, {Vogeley}, \& {Weinberg}}]{blanton03}
{Blanton}, M.~R., {et~al.} 2003, \apj, 594, 186

\bibitem[{{Bower} {et~al.}(1992){Bower}, {Lucey}, \& {Ellis}}]{bower92}
{Bower}, R.~G., {Lucey}, J.~R., \& {Ellis}, R.~S. 1992, \mnras, 254, 601

\bibitem[{{Brammer} {et~al.}(2008){Brammer}, {van Dokkum}, \&
  {Coppi}}]{brammer08}
{Brammer}, G.~B., {van Dokkum}, P.~G., \& {Coppi}, P. 2008, \apj, 686, 1503

\bibitem[{{Brammer} {et~al.}(2012){Brammer}, {van Dokkum}, {Franx},
  {Fumagalli}, {Patel}, {Rix}, {Skelton}, {Kriek}, {Nelson}, {Schmidt},
  {Bezanson}, {da Cunha}, {Erb}, {Fan}, {F{\"o}rster Schreiber}, {Illingworth},
  {Labb{\'e}}, {Leja}, {Lundgren}, {Magee}, {Marchesini}, {McCarthy},
  {Momcheva}, {Muzzin}, {Quadri}, {Steidel}, {Tal}, {Wake}, {Whitaker}, \&
  {Williams}}]{brammer12}
{Brammer}, G.~B., {et~al.} 2012, \apjs, 200, 13

\bibitem[{{Bruzual} \& {Charlot}(2003)}]{bruzual03}
{Bruzual}, G., \& {Charlot}, S. 2003, \mnras, 344, 1000

\bibitem[{{Calzetti} {et~al.}(2000){Calzetti}, {Armus}, {Bohlin}, {Kinney},
  {Koornneef}, \& {Storchi-Bergmann}}]{calzetti00}
{Calzetti}, D., {Armus}, L., {Bohlin}, R.~C., {Kinney}, A.~L., {Koornneef}, J.,
  \& {Storchi-Bergmann}, T. 2000, \apj, 533, 682

\bibitem[{{Cardamone} {et~al.}(2010){Cardamone}, {van Dokkum}, {Urry},
  {Taniguchi}, {Gawiser}, {Brammer}, {Taylor}, {Damen}, {Treister}, {Cobb},
  {Bond}, {Schawinski}, {Lira}, {Murayama}, {Saito}, \&
  {Sumikawa}}]{cardamone10}
{Cardamone}, C.~N., {et~al.} 2010, \apjs, 189, 270

\bibitem[{{Carollo} {et~al.}(2013){Carollo}, {Bschorr}, {Renzini}, {Lilly},
  {Capak}, {Cibinel}, {Ilbert}, {Onodera}, {Scoville}, {Cameron}, {Mobasher},
  {Sanders}, \& {Taniguchi}}]{carollo13}
{Carollo}, C.~M., {et~al.} 2013, \apj, 773, 112

\bibitem[{{Chabrier}(2003)}]{chabrier03}
{Chabrier}, G. 2003, \pasp, 115, 763

\bibitem[{{Cimatti} {et~al.}(2012){Cimatti}, {Nipoti}, \&
  {Cassata}}]{cimatti12}
{Cimatti}, A., {Nipoti}, C., \& {Cassata}, P. 2012, \mnras, 422, L62

\bibitem[{{Cimatti} {et~al.}(2004){Cimatti}, {Daddi}, {Renzini}, {Cassata},
  {Vanzella}, {Pozzetti}, {Cristiani}, {Fontana}, {Rodighiero}, {Mignoli}, \&
  {Zamorani}}]{cimatti04}
{Cimatti}, A., {et~al.} 2004, \nat, 430, 184

\bibitem[{{Cimatti} {et~al.}(2008){Cimatti}, {Cassata}, {Pozzetti}, {Kurk},
  {Mignoli}, {Renzini}, {Daddi}, {Bolzonella}, {Brusa}, {Rodighiero},
  {Dickinson}, {Franceschini}, {Zamorani}, {Berta}, {Rosati}, \&
  {Halliday}}]{cimatti08}
---. 2008, \aap, 482, 21

\bibitem[{{Daddi} {et~al.}(2005){Daddi}, {Renzini}, {Pirzkal}, {Cimatti},
  {Malhotra}, {Stiavelli}, {Xu}, {Pasquali}, {Rhoads}, {Brusa}, {di Serego
  Alighieri}, {Ferguson}, {Koekemoer}, {Moustakas}, {Panagia}, \&
  {Windhorst}}]{daddi05}
{Daddi}, E., {et~al.} 2005, \apj, 626, 680

\bibitem[{{Donley} {et~al.}(2012){Donley}, {Koekemoer}, {Brusa}, {Capak},
  {Cardamone}, {Civano}, {Ilbert}, {Impey}, {Kartaltepe}, {Miyaji}, {Salvato},
  {Sanders}, {Trump}, \& {Zamorani}}]{donley12}
{Donley}, J.~L., {et~al.} 2012, \apj, 748, 142

\bibitem[{{Dressler} {et~al.}(2013){Dressler}, {Oemler}, {Poggianti},
  {Gladders}, {Abramson}, \& {Vulcani}}]{dressler13}
{Dressler}, A., {Oemler}, Jr., A., {Poggianti}, B.~M., {Gladders}, M.~D.,
  {Abramson}, L., \& {Vulcani}, B. 2013, \apj, 770, 62

\bibitem[{{Gon{\c c}alves} {et~al.}(2012){Gon{\c c}alves}, {Martin},
  {Men{\'e}ndez-Delmestre}, {Wyder}, \& {Koekemoer}}]{goncalves12}
{Gon{\c c}alves}, T.~S., {Martin}, D.~C., {Men{\'e}ndez-Delmestre}, K.,
  {Wyder}, T.~K., \& {Koekemoer}, A. 2012, \apj, 759, 67

\bibitem[{{Graves} {et~al.}(2007){Graves}, {Faber}, {Schiavon}, \&
  {Yan}}]{graves07}
{Graves}, G.~J., {Faber}, S.~M., {Schiavon}, R.~P., \& {Yan}, R. 2007, \apj,
  671, 243

\bibitem[{{Grogin} {et~al.}(2011){Grogin}, {Kocevski}, {Faber}, {Ferguson},
  {Koekemoer}, {Riess}, {Acquaviva}, {Alexander}, {Almaini}, {Ashby}, {Barden},
  {Bell}, {Bournaud}, {Brown}, {Caputi}, {Casertano}, {Cassata}, {Castellano},
  {Challis}, {Chary}, {Cheung}, {Cirasuolo}, {Conselice}, {Roshan Cooray},
  {Croton}, {Daddi}, {Dahlen}, {Dav{\'e}}, {de Mello}, {Dekel}, {Dickinson},
  {Dolch}, {Donley}, {Dunlop}, {Dutton}, {Elbaz}, {Fazio}, {Filippenko},
  {Finkelstein}, {Fontana}, {Gardner}, {Garnavich}, {Gawiser}, {Giavalisco},
  {Grazian}, {Guo}, {Hathi}, {H{\"a}ussler}, {Hopkins}, {Huang}, {Huang},
  {Jha}, {Kartaltepe}, {Kirshner}, {Koo}, {Lai}, {Lee}, {Li}, {Lotz}, {Lucas},
  {Madau}, {McCarthy}, {McGrath}, {McIntosh}, {McLure}, {Mobasher},
  {Moustakas}, {Mozena}, {Nandra}, {Newman}, {Niemi}, {Noeske}, {Papovich},
  {Pentericci}, {Pope}, {Primack}, {Rajan}, {Ravindranath}, {Reddy}, {Renzini},
  {Rix}, {Robaina}, {Rodney}, {Rosario}, {Rosati}, {Salimbeni}, {Scarlata},
  {Siana}, {Simard}, {Smidt}, {Somerville}, {Spinrad}, {Straughn}, {Strolger},
  {Telford}, {Teplitz}, {Trump}, {van der Wel}, {Villforth}, {Wechsler},
  {Weiner}, {Wiklind}, {Wild}, {Wilson}, {Wuyts}, {Yan}, \& {Yun}}]{grogin11}
{Grogin}, N.~A., {et~al.} 2011, \apjs, 197, 35

\bibitem[{{Hopkins} {et~al.}(2010){Hopkins}, {Bundy}, {Hernquist}, {Wuyts}, \&
  {Cox}}]{hopkins10}
{Hopkins}, P.~F., {Bundy}, K., {Hernquist}, L., {Wuyts}, S., \& {Cox}, T.~J.
  2010, \mnras, 401, 1099

\bibitem[{{Hopkins} {et~al.}(2009){Hopkins}, {Hernquist}, {Cox}, {Keres}, \&
  {Wuyts}}]{hopkins09scalingrel}
{Hopkins}, P.~F., {Hernquist}, L., {Cox}, T.~J., {Keres}, D., \& {Wuyts}, S.
  2009, \apj, 691, 1424

\bibitem[{{Kajisawa} {et~al.}(2011){Kajisawa}, {Ichikawa}, {Tanaka}, {Yamada},
  {Akiyama}, {Suzuki}, {Tokoku}, {Katsuno Uchimoto}, {Konishi}, {Yoshikawa},
  {Nishimura}, {Omata}, {Ouchi}, {Iwata}, {Hamana}, \& {Onodera}}]{kajisawa11}
{Kajisawa}, M., {et~al.} 2011, \pasj, 63, 379

\bibitem[{{Kewley} {et~al.}(2004){Kewley}, {Geller}, \& {Jansen}}]{kewley04}
{Kewley}, L.~J., {Geller}, M.~J., \& {Jansen}, R.~A. 2004, \aj, 127, 2002

\bibitem[{{Koekemoer} {et~al.}(2011){Koekemoer}, {Faber}, {Ferguson}, {Grogin},
  {Kocevski}, {Koo}, {Lai}, {Lotz}, {Lucas}, {McGrath}, {Ogaz}, {Rajan},
  {Riess}, {Rodney}, {Strolger}, {Casertano}, {Castellano}, {Dahlen},
  {Dickinson}, {Dolch}, {Fontana}, {Giavalisco}, {Grazian}, {Guo}, {Hathi},
  {Huang}, {van der Wel}, {Yan}, {Acquaviva}, {Alexander}, {Almaini}, {Ashby},
  {Barden}, {Bell}, {Bournaud}, {Brown}, {Caputi}, {Cassata}, {Challis},
  {Chary}, {Cheung}, {Cirasuolo}, {Conselice}, {Roshan Cooray}, {Croton},
  {Daddi}, {Dav{\'e}}, {de Mello}, {de Ravel}, {Dekel}, {Donley}, {Dunlop},
  {Dutton}, {Elbaz}, {Fazio}, {Filippenko}, {Finkelstein}, {Frazer}, {Gardner},
  {Garnavich}, {Gawiser}, {Gruetzbauch}, {Hartley}, {H{\"a}ussler},
  {Herrington}, {Hopkins}, {Huang}, {Jha}, {Johnson}, {Kartaltepe},
  {Khostovan}, {Kirshner}, {Lani}, {Lee}, {Li}, {Madau}, {McCarthy},
  {McIntosh}, {McLure}, {McPartland}, {Mobasher}, {Moreira}, {Mortlock},
  {Moustakas}, {Mozena}, {Nandra}, {Newman}, {Nielsen}, {Niemi}, {Noeske},
  {Papovich}, {Pentericci}, {Pope}, {Primack}, {Ravindranath}, {Reddy},
  {Renzini}, {Rix}, {Robaina}, {Rosario}, {Rosati}, {Salimbeni}, {Scarlata},
  {Siana}, {Simard}, {Smidt}, {Snyder}, {Somerville}, {Spinrad}, {Straughn},
  {Telford}, {Teplitz}, {Trump}, {Vargas}, {Villforth}, {Wagner}, {Wandro},
  {Wechsler}, {Weiner}, {Wiklind}, {Wild}, {Wilson}, {Wuyts}, \&
  {Yun}}]{koekemoer11}
{Koekemoer}, A.~M., {et~al.} 2011, \apjs, 197, 36

\bibitem[{{Kriek} {et~al.}(2008){Kriek}, {van der Wel}, {van Dokkum}, {Franx},
  \& {Illingworth}}]{kriek08}
{Kriek}, M., {van der Wel}, A., {van Dokkum}, P.~G., {Franx}, M., \&
  {Illingworth}, G.~D. 2008, \apj, 682, 896

\bibitem[{{Kriek} {et~al.}(2009){Kriek}, {van Dokkum}, {Labb{\'e}}, {Franx},
  {Illingworth}, {Marchesini}, \& {Quadri}}]{kriek09}
{Kriek}, M., {van Dokkum}, P.~G., {Labb{\'e}}, I., {Franx}, M., {Illingworth},
  G.~D., {Marchesini}, D., \& {Quadri}, R.~F. 2009, \apj, 700, 221

\bibitem[{{Kriek} {et~al.}(2006){Kriek}, {van Dokkum}, {Franx}, {Quadri},
  {Gawiser}, {Herrera}, {Illingworth}, {Labb{\'e}}, {Lira}, {Marchesini},
  {Rix}, {Rudnick}, {Taylor}, {Toft}, {Urry}, \& {Wuyts}}]{kriek06}
{Kriek}, M., {et~al.} 2006, \apjl, 649, L71

\bibitem[{{Labb{\'e}} {et~al.}(2005){Labb{\'e}}, {Huang}, {Franx}, {Rudnick},
  {Barmby}, {Daddi}, {van Dokkum}, {Fazio}, {Schreiber}, {Moorwood}, {Rix},
  {R{\"o}ttgering}, {Trujillo}, \& {van der Werf}}]{labbe05}
{Labb{\'e}}, I., {et~al.} 2005, \apjl, 624, L81

\bibitem[{{Lemaux} {et~al.}(2010){Lemaux}, {Lubin}, {Shapley}, {Kocevski},
  {Gal}, \& {Squires}}]{lemaux10}
{Lemaux}, B.~C., {Lubin}, L.~M., {Shapley}, A., {Kocevski}, D., {Gal}, R.~R.,
  \& {Squires}, G.~K. 2010, \apj, 716, 970

\bibitem[{{Marchesini} {et~al.}(2014){Marchesini}, {Muzzin}, {Stefanon},
  {Franx}, {Brammer}, {Marsan}, {Vulcani}, {Fynbo}, {Milvang-Jensen}, {Dunlop},
  \& {Buitrago}}]{marchesini14}
{Marchesini}, D., {et~al.} 2014, ArXiv e-prints

\bibitem[{{Muzzin} {et~al.}(2009){Muzzin}, {van Dokkum}, {Franx}, {Marchesini},
  {Kriek}, \& {Labb{\'e}}}]{muzzin09}
{Muzzin}, A., {van Dokkum}, P., {Franx}, M., {Marchesini}, D., {Kriek}, M., \&
  {Labb{\'e}}, I. 2009, \apjl, 706, L188

\bibitem[{{Muzzin} {et~al.}(2013){Muzzin}, {Marchesini}, {Stefanon}, {Franx},
  {McCracken}, {Milvang-Jensen}, {Dunlop}, {Fynbo}, {Le Fevre}, {Brammer}, \&
  {Labbe}}]{muzzin13}
{Muzzin}, A., {et~al.} 2013, ArXiv e-prints

\bibitem[{{Naab} {et~al.}(2009){Naab}, {Johansson}, \& {Ostriker}}]{naab09}
{Naab}, T., {Johansson}, P.~H., \& {Ostriker}, J.~P. 2009, \apjl, 699, L178

\bibitem[{{Nelson} {et~al.}(2014){Nelson}, {van Dokkum}, {Franx}, {Brammer},
  {Momcheva}, {Schreiber}, {da Cunha}, {Tacconi}, {Bezanson}, {Kirkpatrick},
  {Leja}, {Rix}, {Skelton}, {van der Wel}, {Whitaker}, \& {Wuyts}}]{nelson14}
{Nelson}, E., {et~al.} 2014, \nat, 513, 394

\bibitem[{{Newman} {et~al.}(2014){Newman}, {Ellis}, {Andreon}, {Treu},
  {Raichoor}, \& {Trinchieri}}]{newman14}
{Newman}, A.~B., {Ellis}, R.~S., {Andreon}, S., {Treu}, T., {Raichoor}, A., \&
  {Trinchieri}, G. 2014, \apj, 788, 51

\bibitem[{{Newman} {et~al.}(2012){Newman}, {Ellis}, {Bundy}, \&
  {Treu}}]{newman12}
{Newman}, A.~B., {Ellis}, R.~S., {Bundy}, K., \& {Treu}, T. 2012, \apj, 746,
  162

\bibitem[{{Newman} {et~al.}(2010){Newman}, {Ellis}, {Treu}, \&
  {Bundy}}]{newman10}
{Newman}, A.~B., {Ellis}, R.~S., {Treu}, T., \& {Bundy}, K. 2010, \apjl, 717,
  L103

\bibitem[{{Nipoti} {et~al.}(2012){Nipoti}, {Treu}, {Leauthaud}, {Bundy},
  {Newman}, \& {Auger}}]{nipoti12}
{Nipoti}, C., {Treu}, T., {Leauthaud}, A., {Bundy}, K., {Newman}, A.~B., \&
  {Auger}, M.~W. 2012, \mnras, 422, 1714

\bibitem[{{Oke} {et~al.}(1995){Oke}, {Cohen}, {Carr}, {Cromer}, {Dingizian},
  {Harris}, {Labrecque}, {Lucinio}, {Schaal}, {Epps}, \& {Miller}}]{oke95}
{Oke}, J.~B., {et~al.} 1995, \pasp, 107, 375

\bibitem[{{Onodera} {et~al.}(2012){Onodera}, {Renzini}, {Carollo},
  {Cappellari}, {Mancini}, {Strazzullo}, {Daddi}, {Arimoto}, {Gobat}, {Yamada},
  {McCracken}, {Ilbert}, {Capak}, {Cimatti}, {Giavalisco}, {Koekemoer}, {Kong},
  {Lilly}, {Motohara}, {Ohta}, {Sanders}, {Scoville}, {Tamura}, \&
  {Taniguchi}}]{onodera12}
{Onodera}, M., {et~al.} 2012, \apj, 755, 26

\bibitem[{{Patel} {et~al.}(2013){Patel}, {van Dokkum}, {Franx}, {Quadri},
  {Muzzin}, {Marchesini}, {Williams}, {Holden}, \& {Stefanon}}]{patel13}
{Patel}, S.~G., {et~al.} 2013, \apj, 766, 15

\bibitem[{{Poggianti} {et~al.}(2013){Poggianti}, {Moretti}, {Calvi},
  {D'Onofrio}, {Valentinuzzi}, {Fritz}, \&
  {Renzini}}]{poggianti13numberdensity}
{Poggianti}, B.~M., {Moretti}, A., {Calvi}, R., {D'Onofrio}, M.,
  {Valentinuzzi}, T., {Fritz}, J., \& {Renzini}, A. 2013, \apj, 777, 125

\bibitem[{{Posti} {et~al.}(2014){Posti}, {Nipoti}, {Stiavelli}, \&
  {Ciotti}}]{posti14}
{Posti}, L., {Nipoti}, C., {Stiavelli}, M., \& {Ciotti}, L. 2014, \mnras, 440,
  610

\bibitem[{{Schawinski} {et~al.}(2014){Schawinski}, {Urry}, {Simmons},
  {Fortson}, {Kaviraj}, {Keel}, {Lintott}, {Masters}, {Nichol}, {Sarzi},
  {Skibba}, {Treister}, {Willett}, {Wong}, \& {Yi}}]{schawinski14}
{Schawinski}, K., {et~al.} 2014, \mnras, 440, 889

\bibitem[{{Singh} {et~al.}(2013){Singh}, {van de Ven}, {Jahnke}, {Lyubenova},
  {Falc{\'o}n-Barroso}, {Alves}, {Cid Fernandes}, {Galbany},
  {Garc{\'{\i}}a-Benito}, {Husemann}, {Kennicutt}, {Marino}, {M{\'a}rquez},
  {Masegosa}, {Mast}, {Pasquali}, {S{\'a}nchez}, {Walcher}, {Wild}, {Wisotzki},
  \& {Ziegler}}]{singh13}
{Singh}, R., {et~al.} 2013, \aap, 558, A43

\bibitem[{{Skelton} {et~al.}(2014){Skelton}, {Whitaker}, {Momcheva}, {Brammer},
  {van Dokkum}, {Labbe}, {Franx}, {van der Wel}, {Bezanson}, {Da Cunha},
  {Fumagalli}, {Foerster Schreiber}, {Kriek}, {Leja}, {Lundgren}, {Magee},
  {Marchesini}, {Maseda}, {Nelson}, {Oesch}, {Pacifici}, {Patel}, {Price},
  {Rix}, {Tal}, {Wake}, \& {Wuyts}}]{skelton14}
{Skelton}, R.~E., {et~al.} 2014, ArXiv e-prints

\bibitem[{{Speagle} {et~al.}(2014){Speagle}, {Steinhardt}, {Capak}, \&
  {Silverman}}]{speagle14}
{Speagle}, J.~S., {Steinhardt}, C.~L., {Capak}, P.~L., \& {Silverman}, J.~D.
  2014, ArXiv e-prints

\bibitem[{{Szomoru} {et~al.}(2012){Szomoru}, {Franx}, \& {van
  Dokkum}}]{szomoru12}
{Szomoru}, D., {Franx}, M., \& {van Dokkum}, P.~G. 2012, \apj, 749, 121

\bibitem[{{Toft} {et~al.}(2012){Toft}, {Gallazzi}, {Zirm}, {Wold}, {Zibetti},
  {Grillo}, \& {Man}}]{toft12}
{Toft}, S., {Gallazzi}, A., {Zirm}, A., {Wold}, M., {Zibetti}, S., {Grillo},
  C., \& {Man}, A. 2012, \apj, 754, 3

\bibitem[{{Toft} {et~al.}(2014){Toft}, {Smol{\v c}i{\'c}}, {Magnelli}, {Karim},
  {Zirm}, {Michalowski}, {Capak}, {Sheth}, {Schawinski}, {Krogager}, {Wuyts},
  {Sanders}, {Man}, {Lutz}, {Staguhn}, {Berta}, {Mccracken}, {Krpan}, \&
  {Riechers}}]{toft14}
{Toft}, S., {et~al.} 2014, \apj, 782, 68

\bibitem[{{Trujillo} {et~al.}(2006){Trujillo}, {F{\"o}rster Schreiber},
  {Rudnick}, {Barden}, {Franx}, {Rix}, {Caldwell}, {McIntosh}, {Toft},
  {H{\"a}ussler}, {Zirm}, {van Dokkum}, {Labb{\'e}}, {Moorwood},
  {R{\"o}ttgering}, {van der Wel}, {van der Werf}, \& {van
  Starkenburg}}]{trujillo06}
{Trujillo}, I., {et~al.} 2006, \apj, 650, 18

\bibitem[{{van de Sande} {et~al.}(2013){van de Sande}, {Kriek}, {Franx}, {van
  Dokkum}, {Bezanson}, {Bouwens}, {Quadri}, {Rix}, \& {Skelton}}]{vandesande13}
{van de Sande}, J., {et~al.} 2013, \apj, 771, 85

\bibitem[{{van der Wel} {et~al.}(2012){van der Wel}, {Bell}, {H{\"a}ussler},
  {McGrath}, {Chang}, {Guo}, {McIntosh}, {Rix}, {Barden}, {Cheung}, {Faber},
  {Ferguson}, {Galametz}, {Grogin}, {Hartley}, {Kartaltepe}, {Kocevski},
  {Koekemoer}, {Lotz}, {Mozena}, {Peth}, \& {Peng}}]{vanderwel12}
{van der Wel}, A., {et~al.} 2012, \apjs, 203, 24

\bibitem[{{van der Wel} {et~al.}(2014){van der Wel}, {Franx}, {van Dokkum},
  {Skelton}, {Momcheva}, {Whitaker}, {Brammer}, {Bell}, {Rix}, {Wuyts},
  {Ferguson}, {Holden}, {Barro}, {Koekemoer}, {Chang}, {McGrath},
  {H{\"a}ussler}, {Dekel}, {Behroozi}, {Fumagalli}, {Leja}, {Lundgren},
  {Maseda}, {Nelson}, {Wake}, {Patel}, {Labb{\'e}}, {Faber}, {Grogin}, \&
  {Kocevski}}]{vanderwel14}
---. 2014, \apj, 788, 28

\bibitem[{{van Dokkum} {et~al.}(2009){van Dokkum}, {Kriek}, \&
  {Franx}}]{vandokkum09}
{van Dokkum}, P.~G., {Kriek}, M., \& {Franx}, M. 2009, \nat, 460, 717

\bibitem[{{van Dokkum} {et~al.}(2006){van Dokkum}, {Quadri}, {Marchesini},
  {Rudnick}, {Franx}, {Gawiser}, {Herrera}, {Wuyts}, {Lira}, {Labb{\'e}},
  {Maza}, {Illingworth}, {F{\"o}rster Schreiber}, {Kriek}, {Rix}, {Taylor},
  {Toft}, {Webb}, \& {Yi}}]{vandokkum06}
{van Dokkum}, P.~G., {et~al.} 2006, \apjl, 638, L59

\bibitem[{{van Dokkum} {et~al.}(2008){van Dokkum}, {Franx}, {Kriek}, {Holden},
  {Illingworth}, {Magee}, {Bouwens}, {Marchesini}, {Quadri}, {Rudnick},
  {Taylor}, \& {Toft}}]{vandokkum08}
---. 2008, \apjl, 677, L5

\bibitem[{{van Dokkum} {et~al.}(2010){van Dokkum}, {Whitaker}, {Brammer},
  {Franx}, {Kriek}, {Labb{\'e}}, {Marchesini}, {Quadri}, {Bezanson},
  {Illingworth}, {Muzzin}, {Rudnick}, {Tal}, \& {Wake}}]{vandokkum10}
---. 2010, \apj, 709, 1018

\bibitem[{{Wellons} {et~al.}(2014){Wellons}, {Torrey}, {Ma}, {Rodriguez-Gomez},
  {Vogelsberger}, {Kriek}, {van Dokkum}, {Nelson}, {Genel}, {Pillepich},
  {Springel}, {Sijacki}, {Snyder}, {Nelson}, {Sales}, \&
  {Hernquist}}]{wellons14}
{Wellons}, S., {et~al.} 2014, ArXiv e-prints

\bibitem[{{Whitaker} {et~al.}(2012){Whitaker}, {Kriek}, {van Dokkum},
  {Bezanson}, {Brammer}, {Franx}, \& {Labb{\'e}}}]{whitaker12}
{Whitaker}, K.~E., {Kriek}, M., {van Dokkum}, P.~G., {Bezanson}, R., {Brammer},
  G., {Franx}, M., \& {Labb{\'e}}, I. 2012, \apj, 745, 179

\bibitem[{{Whitaker} {et~al.}(2011){Whitaker}, {Labb{\'e}}, {van Dokkum},
  {Brammer}, {Kriek}, {Marchesini}, {Quadri}, {Franx}, {Muzzin}, {Williams},
  {Bezanson}, {Illingworth}, {Lee}, {Lundgren}, {Nelson}, {Rudnick}, {Tal}, \&
  {Wake}}]{whitaker11}
{Whitaker}, K.~E., {et~al.} 2011, \apj, 735, 86

\bibitem[{{Whitaker} {et~al.}(2013){Whitaker}, {van Dokkum}, {Brammer},
  {Momcheva}, {Skelton}, {Franx}, {Kriek}, {Labb{\'e}}, {Fumagalli},
  {Lundgren}, {Nelson}, {Patel}, \& {Rix}}]{whitaker13}
---. 2013, \apjl, 770, L39

\bibitem[{{Williams} {et~al.}(2009){Williams}, {Quadri}, {Franx}, {van Dokkum},
  \& {Labb{\'e}}}]{williams09}
{Williams}, R.~J., {Quadri}, R.~F., {Franx}, M., {van Dokkum}, P., \&
  {Labb{\'e}}, I. 2009, \apj, 691, 1879

\bibitem[{{Woo} {et~al.}(2014){Woo}, {Dekel}, {Faber}, \& {Koo}}]{woo14}
{Woo}, J., {Dekel}, A., {Faber}, S.~M., \& {Koo}, D.~C. 2014, ArXiv e-prints

\bibitem[{{Wuyts} {et~al.}(2007){Wuyts}, {Labb{\'e}}, {Franx}, {Rudnick}, {van
  Dokkum}, {Fazio}, {F{\"o}rster Schreiber}, {Huang}, {Moorwood}, {Rix},
  {R{\"o}ttgering}, \& {van der Werf}}]{wuyts07}
{Wuyts}, S., {et~al.} 2007, \apj, 655, 51

\bibitem[{{Yan} {et~al.}(2006){Yan}, {Newman}, {Faber}, {Konidaris}, {Koo}, \&
  {Davis}}]{yan06}
{Yan}, R., {Newman}, J.~A., {Faber}, S.~M., {Konidaris}, N., {Koo}, D., \&
  {Davis}, M. 2006, \apj, 648, 281

\bibitem[{{Yesuf} {et~al.}(2014){Yesuf}, {Faber}, {Trump}, {Koo}, {Fang},
  {Liu}, {Wild}, \& {Hayward}}]{yesuf14}
{Yesuf}, H.~M., {Faber}, S.~M., {Trump}, J.~R., {Koo}, D.~C., {Fang}, J.~J.,
  {Liu}, F.~S., {Wild}, V., \& {Hayward}, C.~C. 2014, \apj, 792, 84

\end{thebibliography}

\end{document}